\documentclass [useAMS]{mn2e}
\usepackage {graphicx}
\begin{document}

\title[3:2 HFQPO Pairs and Broad Fe K Line]
{Association of the 3:2 HFQPO Pairs with the Broad Fe K Line in XTE
J1550-564 and GRO J1655-40}

\author[D.-X. Wang, Z.-M. Gan, C.-Y. Huang and Y. Li]
{Ding-Xiong WANG$^{1,2}$, Zhao-Ming GAN$^{1}$, Chang-Yin HUANG$^{1}$
and Yang LI$^{1}$\\\\
$1$ Department of Physics, Huazhong University of Science and
Technology,Wuhan, 430074, China\\
$2$ Send offprint requests to: D.-X. Wang(dxwang@hust.edu.cn)}


\onecolumn
\maketitle

\begin{abstract}
Association of the high-frequency quasi-periodic oscillation (HFQPO)
pairs with the broad Fe K line in XTE J1550-564 and GRO J1655-40 is
discussed based on the magnetic coupling (MC) of a rotating black
hole (BH) with its surrounding disc. The 3:2 HFQPO pairs are
interpreted by virtue of the inner and outer hotspots arising from
non-axisymmetric magnetic field, where the inner hotspot is produced
by a torque exerted at the inner edge of the disc, and the outer
hotspot is created by the screw instability of the large-scale
magnetic field. The very steep emissivity index is created
predominantly by the torque exerted at the inner edge of the disc.
It turns out that the 3:2 HFQPO pairs observed in the two sources
can be fitted by tuning several model parameters, such as the BH
spin, and the main features of this model lie in three aspects. (1)
The condition for only one HFQPO is discussed based on the two
mechanisms for producing the 3:2 HFQPO pairs, (2) an explanation is
given for a systematic shift away from disc dominated flux with the
increasing power-law flux as the HFQPO pairs shift from the higher
to lower frequencies, which is consistent with the analysis given by
Remillard et al. (2002), and (3) the BH spin in XTE J1550-564 and
GRO J1655-40 can be estimated by combining the 3:2 HFQPO pairs with
the very steep emissivity index required for fitting the broad Fe K
emission line.
\end{abstract}


\section{INTRODUCTION}

Quasi-periodic oscillations in black hole X-ray binaries (BHXBs)
have become a very active research field since the launch of the
NASA satellite \emph{RXTE} (Bradt et al. 1993). One of the
remarkable features in BHXBs is that the high-frequency
quasi-periodic oscillations (HFQPOs) could appear in pairs with the
puzzling commensurate frequencies, which have been observed in GRO
J1655-40 (450, 300Hz; Remillard et al. 1999; Strohmayer 2001a;
Remillard et al. 2002, hereafter R02), XTE J1550-564 (276, 184,
92Hz; Miller et al. 2001; R02) and GRS 1915+105 (168, 113, 67, 41Hz;
McClintock \& Remillard 2006; Remillard \& McClintock 2006,
hereafter MR06 and RM06, respectively). The above HFQPOs appear in
pairs with the 3:2 ratio of the higher frequency to the lower
frequency (henceforth 3:2 HFQPO pairs), e.g. 276 vs 184Hz and 450 vs
300Hz occur in XTE J1550-564 and GRO J1655-40, respectively, while
two 3:2 HFQPO pairs, 168 vs 113Hz and 67 vs 41Hz, appear in GRS
1915+105.

A number of models have been proposed to explain the origin of HFQPO
pairs in BHXBs. Strohmayer (2001a, 2001b) investigated combinations
of the azimuthal and radial coordinate frequencies in general
relativity to explain the HFQPO pairs in GRO J1655-40 and GRS
1915+105. Wagoner et al. (2001) regarded the HFQPO pairs as
fundamental g-mode and c-mode discoseismic oscillations in a
relativistic accretion disc. Very recently, Silbergleit \& Wagoner
(2007) discussed the corotation resonance and diskoseismology modes
of black hole (BH) accretion discs and suggested that the HFQPO
pairs may relate to the excitation of two (groups of) g-modes of
discoseismic oscillations. Abramowicz \& Kluzniak (2001) explained
the pairs in GRO J1655-40 as a resonance between orbital and
epicyclic motion of accreting matter. Recently, the resonance model
is presented in a more realistic context, in which "parametric
resonance" concept is introduced to describe the oscillations rooted
in fluid flow where there is a coupling between the radial and polar
coordinate frequencies (Abramowicz et al. 2003; Kluzniak et al.
2004; T\"{o}r\"{o}k et al. 2005). As argued by van der Klis (2000,
2006), HFQPOs in X-ray binaries probably originate from the inner
edge of an accretion disc around a neutron star or a stellar-mass
BH, since millisecond is the natural timescale for accretion process
in these regions.

On the other hand, Wilms et al. (2001) presented the first
XMM-Newton observation of MCG-6-30-15, and they found a broad Fe
disc line from the observed spectrum. However, no quasi-periodic
oscillations have been detected in this source. A very steep
emissivity index ($4.3<\beta<5.0$) from the inner accretion disc is
required for fitting the broad Fe K emission line, which is
difficult to be explained within the framework of a standard
accretion disc (SAD). Wilms et al. (2001) suggested that the
magnetic extraction of rotational energy from a spinning black hole
should be invoked to create this steep emissivity.

The relativistic disc lines for microquasars have been analyzed by
some authors, e.g., XTE J1550-564 by Sobczak et al. (2000), Miller
et al. (2003, 2004) ; XTE J1650-500 by Miller et al. (2004),
Miniutti, Fabian, and Miller (2004); GRO J1655-40 by Miller et al.
(2004), Diaz Trigo et al. (2007). However, the steep emissivity
indexes required by broad Fe K $\alpha$ lines were hardly worked out
except a few cases, such as, $\beta =5.5^{+0.5}_{-0.7}$ for GX 339-4
in the very high state, and $\beta \simeq 5$ for XTE J1650-500 based
on the observation of XMM-Newton. A detailed review for relativistic
X-ray emission lines from the inner accretion disc around black
holes for AGNs and stellar-mass black holes are given by Miller
(2007).

With the existence of a magnetic field connecting a rotating black
hole to its surrounding disc, energy and angular momentum can be
transferred from the black hole to the disc, and this energy
mechanism is referred to as the MC process, which is regarded as one
of the variants of the Blandford-Znajek (BZ) process (Blandford {\&}
Znajek 1977, 1999; Li 2000, 2002, hereafter L02, Wang 2002).

Wang et al. (2002, 2003a, hereafter W03a) incorporated the BZ and MC
processes into black hole accretion disc, and henceforth this model
is referred to as MC-I model, which was used to interpret the very
steep  emissivity index required by the observation of MCG-6-30-15.
In addition, Wang et al. (2005, hereafter W05) suggested that the
3:2 HFQPO pairs observed in the above BHXBs can be fitted by virtue
of two hotspots in the inner disc, which are produced by the
non-axisymmetric MC process.

The fatal shortcoming of MC-I model is that the fits of the 3:2 HFQPO pairs
always accompany jets, which is driven by the BZ process. As reviewed in
MR06 and RM06, jets are generally observed in the hard state of BHXBs, while
HFQPOs are always detected in the steep power-law (SPL) state. Thus the fits
of the 3:2 HFQPO pairs based on MC-I model are inconsistent with the
observations.

Based on a careful analysis of the 1998-1999 outburst of XTE
J1550-564 and the 1996-1997 outburst of GRO J1655-40 some
interesting features have been found in R02, the 3:2 HFQPO pairs
could be detected simultaneously in the SPL state, and a systematic
shift away from disc dominated flux with the increasing power-law
flux is detected as the HFQPO pairs shift from 276 to 184 Hz for XTE
J1550-564 and from 450 to 300 Hz for GRO J1655-40 as shown in Figs.
5 and 7 of R02, respectively. In addition, Remillard (2004) noted
that the broad Fe K emission lines can be also detected in the SPL
state in XTE J1550-564 and GRO J1655-40. These results imply that
the Fe K lines could be detected simultaneously with the 3:2 HFQPO
pairs in these two sources.\\

In this paper, we interpret the 3:2 HFQPO pairs associated with the
broad Fe K line in XTE J1550-564 and GRO J1655-40 by modifying MC-I
model in two aspects. (1) The open field lines corresponding to the
BZ process is removed, and (2) a non-zero torque exerted at the
inner edge of the disc is introduced. Henceforth the modified model
is referred to as MC-II model. It turns out that the 3:2 HFQPO pairs
observed in the two sources can be fitted by tuning several model
parameters, such as the BH spin, and the main features of MC-II
model lie in the following aspects: (1) The condition for only one
HFQPO is discussed based on the two mechanisms for producing the 3:2
HFQPO pairs. (2) An explanation is given for a systematic shift away
from disc dominated flux with the increasing power-law flux as the
HFQPO pairs shift from the higher to lower frequencies, which is
consistent with the analysis given by R02. (3) The BH spin in XTE
J1550-564 and GRO J1655-40 can be estimated by combining the 3:2
HFQPO pairs with the very steep emissivity index required for
fitting the broad Fe K emission line.

This paper is organized as follows. In \S 2 we present a detailed
description of MC-II model, in which the main differences between
MC-I and MC-II models are outlined. In \S 3 we describe the elements
for fitting the 3:2 HFQPO pairs observed in XTE J1550-564 and GRO
J1655-40, and present the calculation sequence and results based on
MC-II model. We discuss the possible correlation of the 3:2 HFQPO
pairs with the very steep emissivity index required by broad Fe K
$\alpha$ lines. It is expected that the 3:2 HFQPO pairs could be
observed with the steep emissivity index, and the black hole spin
could be estimated once the value ranges of the index are determined
in the future observations. Finally, we discuss some issues related
to MC-II model in \S 4.

Throughout this paper the geometric units $G = c = 1$ are used.

\section{DESCRIPTION OF MC-II MODEL }

In this section we present a detailed description of MC-II model based on
the following assumptions.

(1) The accretion disc is perfectly conducting, and the closed
magnetic field lines are frozen in the disc. The disc is thin and
relativistic, lying in the equatorial plane of a Kerr black hole.

(2) Corona is introduced, which consists of tenuous hot plasma above
the disc surface as shown in Figure 1. Both large- and small-scale
magnetic fields are involved in MC-II model, in which the
large-scale magnetic field plays a key role in transferring energy
and angular momentum from a fast-rotating black hole to its
surrounding disc in the MC process, while the small-scale magnetic
field energizes the corona above the disc by virtue of some physical
processes, such as magnetic reconnection and buoyancy (Haardt {\&}
Maraschi 1991). It has been shown that the existence of corona can
improve the fitting of the output spectra from the BHXBs, which is
helpful to interpret the X-ray radiation observed in the SPL state
(Ma et al. 2006).


(3) The large-scale magnetic field is assumed to be non-axisymmetric
as described by Wang et al. (2003b, hereafter W03b). In addition,
the magnetic field is assumed to be constant on the black hole
horizon, and it varies as a power law with disc radius, i.e., $B_D
\propto r^{ - n}$.

(4) One of the main differences between MC-I and MC-II models lies
in the fact that ``no torque boundary condition'' is assumed in MC-I
model as given in SAD (Novikov {\&} Thorne 1973, hereafter NT73;
Page {\&} Thorne 1974), while a nonzero torque is exerted at the
inner edge of the disc in MC-II model. This torque arises from the
magnetic connection of the matter in the plunging region with the
matter in the inner disc as argued by some authors (Livio, Ogilivie
{\&} Pringle 1999, hereafter L99; Krolik 1999; Gammie 1999; Agol
{\&} Krolik 2000).

(5) Another difference between the two models lies in the fact that
the open field lines are removed from MC-II model, while the closed
field lines connecting the black hole with its surrounding accretion
disc are retained, as shown in Figure 1. So only the MC process
rather than the BZ process can work in MC-II model.

The reason for the modification of the magnetic field configurations lies in
two points.

(i) The existence of the closed field lines depends upon the mapping
relation with the conservation of the magnetic flux as argued in
W03a, being not the sufficient condition for the existence of the
open field lines at the horizon.

(ii) The second reason is more important, i.e., the 3:2 HFQPO pairs
are observed in the SPL state, in which no jets are observed. Thus
the magnetic field configuration depicted in Figure 1 is more
consistent with the observations.\\

In order to illustrate the elements for interpreting the 3:2 HFQPO
pairs we derive the radiation flux from the accretion disc based on
MC-II model. According to the conservation of energy and angular
momentum we have the following relations given in L02 and W03a,

\begin{equation}
\label{eq1}
\left\{ {\begin{array}{l}
 \frac{d}{dr}(\dot {M}_D L^\dag - g_{vis} ) = 4\pi r(F_{total} L^\dag -
H_{MC} ),\mbox{ } \\\\
 \frac{d}{dr}(\dot {M}_D E^\dag - g_{vis} \Omega _D ) = 4\pi r(F_{total}
E^\dag - H_{MC} \Omega _D ). \\
 \end{array}} \right.
\end{equation}

In equation (\ref{eq1}) $\dot {M}_D $ is the accretion rate. The
quantities $g_{vis} $ and $H_{MC} $ are respectively the interior
viscous torque in the disc and the flux of angular momentum
transferred between the black hole and the disc in the MC process,
and $E^\dag $ and $L^\dag $ are the specific energy and angular
momentum of the disc matter, being given in NT73 as follows (where
$\chi \equiv \sqrt {r / M} $),

\begin{equation}
\label{eq8} E^\dag = {\left( {1 - 2\chi ^{ - 2} + a_ * \chi ^{ - 3}}
\right)} \mathord{\left/ {\vphantom {{\left( {1 - 2\chi ^{ - 2} + a_
* \chi ^{ - 3}} \right)} {\left( {1 - 3\chi ^{ - 2} + 2a_ * \chi ^{
- 3}} \right)^{1 / 2}}}} \right. \kern-\nulldelimiterspace} {\left(
{1 - 3\chi ^{ - 2} + 2a_ * \chi ^{ - 3}} \right)^{1 / 2}},
\end{equation}

\begin{equation}
\label{eq9} L^\dag = M\chi {\left( {1 - 2a_ * \chi ^{ - 3} + a_ * ^2
\chi ^{ - 4}} \right)} \mathord{\left/ {\vphantom {{\left( {1 - 2a_
* \chi ^{ - 3} + a_ * ^2 \chi ^{ - 4}} \right)} {\left( {1 - 3\chi
^{ - 2} + 2a_ * \chi ^{ - 3}} \right)^{1 / 2}}}} \right.
\kern-\nulldelimiterspace} {\left( {1 - 3\chi ^{ - 2} + 2a_ * \chi
^{ - 3}} \right)^{1 / 2}}.
\end{equation}

By resolving equation (\ref{eq1}) we derive the total radiation flux
$F_{total} $ and the interior viscous torque $g_{vis} $ in the disc
as follows,

\begin{equation}
\label{eq2}
F_{total} = F_{DA} + F_{MC} + F_{gin} ,
\end{equation}

\begin{equation}
\label{eq3}
g_{vis} = \frac{E^\dag - \Omega _D L^\dag }{ - d\Omega _D / dr}4\pi
rF_{total} .
\end{equation}

In equation (\ref{eq2}) $F_{DA} $ and $F_{MC} $ are respectively the
radiation fluxes contributed by the disc accretion and the MC
process, and $F_{gin} $ is the extra contribution due to the
exterior torque exerted at the inner edge, and $\Omega _D $ is the
angular velocity of the relativistic thin disc and it reads

\begin{equation}
\label{eq4}
\Omega _D = \frac{1}{M\left( {\chi ^3 + a_ * } \right)}.
\end{equation}

In equation (\ref{eq4}) the parameter $a_ * \equiv J \mathord{\left/
{\vphantom {J {M^2}}} \right. \kern-\nulldelimiterspace} {M^2}$ is
the black hole spin, which is defined in terms of the black hole
mass $M$ and angular momentum $J$, and the parameter $\chi $ is
related to the disc radius by $\chi = \sqrt \xi \chi _{ms} $, where
$\xi \equiv r \mathord{\left/ {\vphantom {r {r_{ms} }}} \right.
\kern-\nulldelimiterspace} {r_{ms} }$ is defined in terms of the
innermost stable circular orbit (ISCO, NT73). The three parts of
$F_{total} $ are expressed as follows,

\begin{equation}
\label{eq5}
F_{DA} = \frac{1}{4\pi r}\frac{ - {d\Omega _D } \mathord{\left/ {\vphantom
{{d\Omega _D } {dr}}} \right. \kern-\nulldelimiterspace} {dr}}{\left(
{E^\dag - \Omega _D L^\dag } \right)^2}\int_{r_{in} }^r {\left( {E^\dag -
\Omega _D L^\dag } \right)} \left( {\dot {M}_D {dL^\dag } \mathord{\left/
{\vphantom {{dL^\dag } {dr}}} \right. \kern-\nulldelimiterspace} {dr}}
\right)dr,
\end{equation}

\begin{equation}
\label{eq6}
F_{MC} = - \frac{d\Omega _D }{rdr}\left( {E^\dag - \Omega _D L^\dag }
\right)^{ - 2}\int_{r_{ms} }^r {\left( {E^\dag - \Omega _D L^\dag }
\right)H_{MC} rdr} ,
\end{equation}

\begin{equation}
\label{eq7} F_{gin} = \frac{1}{4\pi r}\frac{ - d\Omega _D /
dr}{(E^\dag - \Omega _D L^\dag )^2}\cdot g_{in} \left( {E_{in}^\dag
- \Omega _{in} L_{in}^\dag } \right),
\end{equation}

\noindent
where $\Omega _{in} = \Omega _D (r_{in} )$ is the angular velocity at
$r_{in} $, other terms with subscripts ``\textit{in}'' are taken their values at
$r_{in} $.

The function $H_{MC} $ in equation (\ref{eq6}) is related to the MC
torque by ${\partial T_{MC}^{NA} } \mathord{\left/ {\vphantom
{{\partial T_{MC}^{NA} } {\partial r}}} \right.
\kern-\nulldelimiterspace} {\partial r} = 4\pi rH_{MC} $, and
$T_{MC}^{NA} $ is the MC torque due to non-axisymmetric magnetic
field and it reads (W03b)

\begin{equation}
\label{eq10}
T_{MC}^{NA} = \lambda T_{MC}^A ,
\end{equation}

\noindent
where $T_{MC}^A $ is the MC torque due to axisymmetric magnetic field, being
expressed in W03a as

\begin{equation}
\label{eq11}
{T_{MC}^A } \mathord{\left/ {\vphantom {{T_{MC}^A } {T_0 }}} \right.
\kern-\nulldelimiterspace} {T_0 } = 4a_ * \left( {1 + q} \right)\int_{\theta
_S }^\theta {\frac{\left( {1 - \beta _\Omega } \right)\sin ^3\theta d\theta
}{2 - \left( {1 - q} \right)\sin ^2\theta }} ,
\quad
\theta _S < \theta < \theta _L .
\end{equation}

The parameter $\lambda $ in equation (\ref{eq10}) is used to
indicate the difference between $T_{MC}^A $ and $T_{MC}^{NA} $ as
described in W03b, and $\beta _\Omega \equiv {\Omega _D }
\mathord{\left/ {\vphantom {{\Omega _D } {\Omega _H }}} \right.
\kern-\nulldelimiterspace} {\Omega _H }$ is the ratio of the angular
velocity of the disc to that of the black hole. It is easy to check
by using equation (\ref{eq4}) that $\beta _\Omega < 1$ always holds
for a fast-spinning black hole with $a_ * > 0.3594$, implying that
energy and angular momentum are transferred from the black hole to
the inner disc in the MC process.

Since the magnetic field on the black hole is supported by the
surrounding disc, there is some relation between $B_H $ and $\dot
{M}_D $. One possibility has been suggested by Moderski et al.
(1997), which is based upon the balance between the pressure of the
magnetic field on the horizon and the ram pressure of the innermost
parts of an accretion flow, i.e.,

\begin{equation}
\label{eq12}
{B_H^2 } \mathord{\left/ {\vphantom {{B_H^2 } {\left( {8\pi } \right)}}}
\right. \kern-\nulldelimiterspace} {\left( {8\pi } \right)} = P_{ram} \sim
\rho c^2\sim {\dot {M}_D } \mathord{\left/ {\vphantom {{\dot {M}_D } {\left(
{4\pi r_H^2 } \right)}}} \right. \kern-\nulldelimiterspace} {\left( {4\pi
r_H^2 } \right)},
\end{equation}

\noindent
where $r_H $ is the radius of the black hole horizon. Considering that
equation (\ref{eq12}) is not a certain relation between $B_H $ and $\dot {M}_D $, we
rewrite it as follows,

\begin{equation}
\label{eq13}
\dot {M}_D = \alpha _m B_H^2 r_H^2 ,
\end{equation}

\noindent
where $\alpha _m $ is a parameter to adjust the accretion rate $\dot {M}_D
$.

Inspecting equation (\ref{eq7}), we find that the contribution
$F_{gin} $ is related directly to the exterior torque $g_{in} $.
According to L99 the strength of the magnetic field produced by
dynamo process in the disc is given by

\begin{equation}
\label{eq14}
\frac{B_{dyn}^2 }{4\pi } \sim \frac{W}{2h_{in} } = w_\varphi ^r ,
\end{equation}

\noindent
and the torque $g_{in} $ is related to the viscous stress $w_\varphi ^r $ at
$r_{in} $ as follows,

\begin{equation}
\label{eq15}
g_{in} = r_{in} \cdot 2\pi r_{in} \cdot 2h_{in} \cdot w_\varphi ^r ,
\end{equation}

\noindent
where $h_{in} $ is the half-thickness at $r_{in} $, being related to $r_{in}
$ by $h_{in} = \alpha _H r_{in} $.

In equation (\ref{eq14}) $W$ is the integrated shear stress of the
disc, and the disc magnetic field $B_D $ is related to $B_{dyn} $ by

\begin{equation}
\label{eq16}
B_D \sim \left( {h \mathord{\left/ {\vphantom {h r}} \right.
\kern-\nulldelimiterspace} r} \right)_{\max } B_{dyn} ,
\end{equation}

\noindent where $h$ is the half-thickness of the disc. As argued in
W03a, the field $B_D $ is related to $B_H $ by

\begin{equation}
\label{eq17}
B_D = B_H \left( {{r_H } \mathord{\left/ {\vphantom {{r_H } {\varpi _{in}
}}} \right. \kern-\nulldelimiterspace} {\varpi _{in} }} \right),
\end{equation}

\noindent
where $\varpi _{in} $ is the cylindrical radius at $r_{in} $ in the context
of Kerr metric.

Incorporating equations (\ref{eq16}) and (\ref{eq17}), we have

\begin{equation}
\label{eq18}
B_{dyn}^2 = \left( {{r_{in} } \mathord{\left/ {\vphantom {{r_{in} } {h_{in}
}}} \right. \kern-\nulldelimiterspace} {h_{in} }} \right)^2B_D^2 = \left[
{{\left( {r_{in} r_H } \right)} \mathord{\left/ {\vphantom {{\left( {r_{in}
r_H } \right)} {\left( {\varpi _{in} h_{in} } \right)}}} \right.
\kern-\nulldelimiterspace} {\left( {\varpi _{in} h_{in} } \right)}}
\right]^2B_H^2 .
\end{equation}

Incorporating equations (\ref{eq14}), (\ref{eq15}) and (\ref{eq18}), we have the torque $g_{in} $ as
follows,

\begin{equation}
\label{eq19}
g_{in} = \alpha _H B_{dyn}^2 r_{in}^3 = \alpha _H^{ - 1} r_{in}^3 \left(
{{r_H } \mathord{\left/ {\vphantom {{r_H } {\varpi _{in} }}} \right.
\kern-\nulldelimiterspace} {\varpi _{in} }} \right)^2B_H^2 .
\end{equation}

Assume that the inner edge of the disc is initially located at ISCO.
If the magnetic pressure inside ISCO is strong enough, a radial
force can be exerted at the inner edge of the disc due to the
magnetic connection, resulting in an outward displacement of the
inner edge from $r_{ms} $ to $r_{in} $. It is assumed that the
magnetic torque $g_{in} $ exerted at $r_{in} $ is determined by the
following equations,

\begin{equation}
\label{eq20}
\delta g_{in} \Omega _{in} = \left[ {E^\dag \left( {r_{in} ,a_ * } \right) -
E^\dag \left( {r_{ms} ,a_ * } \right)} \right]\dot {M}_D ,
\end{equation}

\noindent
where $g_{in} \Omega _{in} $ is the power of the magnetic torque, and
$\delta $ is the fraction of the power keeping the inner edge at $r_{in} $.

\section{FITTING 3:2 HFQPO PAIRS ASSOCIATED WITH BROAD FE K LINE}

Considering the fact that the accretion disc is perfectly
conducting, and the closed magnetic field lines are frozen in the
disc, and the inner and outer hotspots are produced by
non-axisymmetric magnetic field, we infer that these hotspots rotate
with the angular velocity of the disc, and the upper and lower
frequencies of the 3:2 HFQPO pairs can be expressed by

\begin{equation}
\label{eq21}
\nu _{HFQPO} = \nu _0 \left( {\xi ^{3 \mathord{\left/ {\vphantom {3 2}}
\right. \kern-\nulldelimiterspace} 2}\chi _{ms}^3 + a_ * } \right)^{ - 1},
\end{equation}

\noindent
where $\nu _0 \equiv \left( {m_{BH} } \right)^{ - 1}\times 3.23\times
10^4Hz$, and $m_{BH} \equiv M \mathord{\left/ {\vphantom {M {M_ \odot }}}
\right. \kern-\nulldelimiterspace} {M_ \odot }$ is the black hole mass in
terms of one solar mass. Equation (\ref{eq21}) can be derived directly from equation
(\ref{eq4}).

The lower frequency $\nu _{lower} $ of the HFQPO pair is produced by the
outer hotspot, which is located at the outer boundary of the MC region as
shown in Figure 1. According to the Kruskal-Shafranov criterion, the screw
instability of the large-scale magnetic field will occur, if the magnetic
field line turns around itself about once (Kadomtsev 1966; Bateman 1978).
Based on the Kruskal-Shafranov criterion the radius $r_{out} $ can be
determined by the following equation (Wang et al. 2004),

\begin{equation}
\label{eq22}
{\left( {{2\pi \varpi _D } \mathord{\left/ {\vphantom {{2\pi \varpi _D } L}}
\right. \kern-\nulldelimiterspace} L} \right)B_D^p } \mathord{\left/
{\vphantom {{\left( {{2\pi \varpi _D } \mathord{\left/ {\vphantom {{2\pi
\varpi _D } L}} \right. \kern-\nulldelimiterspace} L} \right)B_D^p } {B_D^T
}}} \right. \kern-\nulldelimiterspace} {B_D^T } \le 1,
\end{equation}

\noindent where $B_D^p $ and $B_D^T $ are the poloidal and toroidal
components of the magnetic field on the disc, respectively. The
quantities $L$ and $\varpi _D $ are respectively the poloidal length
of the field line and the cylindrical radius on the disc, and the
latter reads

\begin{equation}
\label{eq23} \varpi _D = {\Sigma _D } \mathord{\left/ {\vphantom
{{\Sigma _D } {\rho _D }}} \right. \kern-\nulldelimiterspace} {\rho
_D } = \xi M\chi _{ms}^2 \sqrt {1 + a_ * ^2 \xi ^{ - 2}\chi _{ms}^{
- 4} + 2a_ * ^2 \xi ^{ - 3}\chi _{ms}^{ - 6} } ,
\end{equation}

\noindent where $\Sigma _D $ and $\rho _D $ are the Kerr metric
parameters (Novikov {\&} Thorns 1973).

Wang et al. (2004) expressed equation (\ref{eq22}) in terms of the parameters
involved in the MC process as follows,

\begin{equation}
\label{eq24}
\left( {{2\pi \varpi _D } \mathord{\left/ {\vphantom {{2\pi \varpi _D } L}}
\right. \kern-\nulldelimiterspace} L} \right)F_{SC} \left( {a_ * ,\xi ,n}
\right) \le 1.
\end{equation}

The equality in equation (\ref{eq24}) corresponds to the screw
instability, and $F_{SC}( {a_ * ,\xi ,n}) $ is a function of the
parameters, $a_ * $, $\xi $ and $n$, and it reads

\begin{equation}
\label{eq25}
F_{SC}( {a_ * ,\xi ,n}) = \frac{\xi ^{1 - n}\left( {1 +
q} \right)\left[ {2\csc ^2\theta - \left( {1 - q} \right)}
\right]}{2a_ * \left( {1 - \beta _\Omega } \right)}\sqrt {\frac{1 +
a_ * ^2 \chi _{ms}^{ - 4} \xi ^{ - 2} + 2a_ * ^2 \chi _{ms}^{ - 6}
\xi ^{ - 3}}{1 + a_ * ^2 \chi _{ms}^{ - 4} + 2a_ * ^2 \chi _{ms}^{ -
6} }} ,
\end{equation}

\noindent
where $q \equiv \sqrt {1 - a_ * ^2 } $ is a function of the black hole spin.
Thus the position of the outer hotspot can be determined by the criterion
(\ref{eq24}) for the given $a_ * $ and $n$.

The upper frequency $\nu _{higher} $ of the HFQPO pair is produced
by the inner hotspot, of which the position is determined by the
maximum of the following function,

\begin{equation}
\label{eq26}
F_{HFQPO} \equiv {\xi ^2F_{total} } \mathord{\left/ {\vphantom {{\xi
^2F_{total} } {F_0 }}} \right. \kern-\nulldelimiterspace} {F_0 },
\end{equation}

\noindent
where $F_0 \equiv B_H^2 = B_8^2 \times \left( {3\times 10^{26}erg \cdot s^{
- 1} \cdot cm^{ - 2}} \right)$ is defined as a unit of the
radiation flux, and $B_8 $ is the magnetic field in terms
of $10^8gauss$.

The emissivity index can be calculated from the total radiation flux
$F_{total} $ based on the definition as follows,

\begin{equation}
\label{eq27}
\beta = - {d\ln F_{total} } \mathord{\left/ {\vphantom {{d\ln F_{total} }
{d\ln r}}} \right. \kern-\nulldelimiterspace} {d\ln r}.
\end{equation}

Incorporating equations (\ref{eq2}), (\ref{eq5}), (\ref{eq6}), (\ref{eq7}), (\ref{eq19}) and (\ref{eq27}), we obtain the
emissivity index $\beta = \beta _{DA} + \beta _{MC} + \beta _{gin} $ by
calculating the following contributions,

\begin{equation}
\label{eq28}
\beta _{DA} \equiv - \frac{r}{F_{total} }\frac{dF_{DA} }{dr},
\quad
\beta _{MC} \equiv - \frac{r}{F_{total} }\frac{dF_{MC} }{dr},
\quad
\beta _{gin} \equiv - \frac{r}{F_{total} }\frac{dF_{gin} }{dr},
\end{equation}

\noindent where $\beta _{DA} $, $\beta _{MC} $ and $\beta _{gin} $
are contributed by the disc accretion, the MC process and the
magnetic torque exerted at the inner edge of the disc, respectively.

Based on equations (\ref{eq5}), (\ref{eq6}), (\ref{eq7}) and
(\ref{eq28}) we obtain the curves of the above contributions to the
total radiation flux and the emissivity index versus the disc radius
for XTE J1550-564 with $m_{BH} = \mbox{10.8}$ as shown in Figure 2a
and 2b, respectively.


As shown in Figure 2 the contributions of the disc accretion and the
MC process to the total radiation flux and the emissivity index are
more than two orders of magnitude less than those due to the
magnetic torque at the inner edge of the disc, and both the total
radiation flux $F_{total} $ and the emissivity index $\beta $ take
their maxima at $r_{in} $. Thus we infer that both the inner hotspot
and the maximum emissivity index are located at $r_{in} $.

Combining equation (\ref{eq21}) with the higher and lower HFQPO
frequencies corresponding to the disc radii  $r_{in}=\xi_{in}
r_{ms}$ and $r_{out}=\xi_{out} r_{ms}$, we have

\begin{equation}
\label{eq29} \nu_{higher}=\nu_{0} (\xi^{3/2}_{in} \chi^{3}_{ms}+a_*
)^{-1}
\end{equation}

\begin{equation}
\label{eq30}\nu_{lower}=\nu_0 (\xi^{3/2}_{out} \chi^3_{ms}+a_*
)^{-1}
\end{equation}

Equations (\ref{eq20}),(\ref{eq24}),(\ref{eq27}),(\ref{eq29}) and
(\ref{eq30}) can be regarded as a set of equations for calculating
the disc radii, $r_{in}$  and $r_{out}$ , for the 3:2 HFQPO pairs,
and henceforth \emph{the five independent equations} are referred to
as \textbf{FIE}. Four input parameters, $a_ * $, $m_{BH} $, $\nu
_{higher} $ and $\nu _{lower} $, are involved, and five output
parameters, $n$, $r_{in} $, $r_{out} $, $\delta $ and $\beta $ are
obtained based on \textbf{FIE}. The detailed calculation sequence is
described as follows.

(1) As the first step, we assume a value of spin with the observed
black hole mass, e.g., $a_ * = 0.5$ and $m_{BH} = \mbox{10.8}$ for
XTE J1550-564. Based on the higher frequency $\nu _{higher} $, e.g.,
276Hz for XTE J1550-564, the position of the inner hotspot can be
determined by equation (\ref{eq29}). Thus the radius $r_{in} $ can
be determined for the given black hole spin and mass.

(2) In the second step, the position of the outer hotspot
corresponding to the lower frequency $\nu_{lower}$ , e.g., 184Hz for
XTE J1550-564, can be determined by combining equation (\ref{eq30})
with the criterion of the screw instability given by equations
(\ref{eq24}), and the power law index $n$ indicating the variation
of the magnetic field with the disc radius can be determined in
calculating $\nu_{lower}$ by virtue of the criterion of the screw
instability.

(3) Finally, in the third step, the outward displacement of the
inner edge from $r_{ms}$ to $r_{in}$ can be worked out, and the
parameter $\delta$ is determined by equation (\ref{eq20}). Combining
equation (\ref{eq27}) with equation (\ref{eq2}) and the obtained
values $a_*$, $r_{in}$ and $n$, we have a maximum value of the
emissivity index $\beta$ located at $r_{in}$.

In this way, we can obtain one set of solution of $r_{in} $,
$r_{out} $, $n$, $\delta $ and $\beta $ by inputting each black hole
spin $a_ * $ with the observed $\nu _{higher} $, $\nu _{lower} $ and
$m_{BH} $ into \textbf{FIE}, and the curves of $\beta $, $n$ and
$\delta $ varying with $a_ * $ for XTE J1550-564 and GRO J1655-40
are shown in Figures 3 and 4, respectively.



By resolving \textbf{FIE} we have the specific solutions
corresponding to the two BHXBs with the lower and higher black hole
masses and different spins as listed in Table 1.


Inspecting Figures 3 and 4, and Table 1, we find that the 3:2 HFQPO
pairs observed in XTE J1550-564 and GRO J1655-40 can be worked out
with the steep emissivity index, and the features of MC-II model are
summarized as follows.

(1) For a given black hole mass, a bigger spin corresponds to a
smaller emissivity index $\beta $ and a smaller parameter $n$, but
to a bigger parameter $\delta $. For a given black hole spin, a
bigger black hole mass corresponds to a bigger $\beta $, while the
parameters $n$ and \textit{$\delta $ }are insensitive to the
variation of the black hole mass. This result can be understood
based on black hole physics (Shapiro \& Teukolsky 1983). The width
of the plunging region decreases monotonically with the increasing
black hole spin, so the effects of the magnetic torque in the
plunging region decreases with the increasing black hole spin.

(2) The black hole spin could be estimated, if the value range of
the emissivity index can be determined by the future observations.
As shown by the shaded regions in Figures 3a and 4a, the black hole
spin could be constrained in the range of $0.7 < a_ * < 0.998$ with
the emissivity indexes in the ranges of $3.95 < \beta < 4.12$ and
$3.95 < \beta < 4.24$ for XTE J1650-564 and GRO J1655-40,
respectively.

As pointed out in RM06, HFQPOs are likely to offer the most reliable
measurement of BH spins once the correct model is known. The
observations involved HFQPOs are very complicated, behaving multiple
features in different sources, and a variety of mechanisms and
models are invoked to fit the HFQPOs in BHXBs. The model of the MC
process with a nonzero torque exerted at the inner edge of accretion
disc provides an interpretation for the association of the 3:2 HFQPO
pairs with the broad Fe K line, and it could be a possible approach
to the BH spins of the binaries in the Galaxy.

\section{DISCUSSION}

In this paper, association of the 3:2 HFQPO pairs with the broad Fe
K line in XTE J1550-564 and GRO J1655-40 is discussed based on MC-II
model. It is shown that the 3:2 HFQPO pairs observed in the two
sources can be fitted with the observational constraints by tuning
several model parameters. Some related issues are given as follows.

(1) It is noted that the mechanisms of creating the two hotspots are
different: the inner one is produced predominantly by the magnetic
torque exerted at the inner edge of the disc, while the outer one
arises from the screw instability of the large-scale magnetic field.
It is helpful to imagine the magnetic field line as an elastic
string. The rotating BH twists the field line, while the field line
tries to untwist itself. Once the toroidal component of the magnetic
field is strong enough to satisfy the criterion, the screw
instability will occurs, just as a twisted elastic string releases
its energy under appropriate conditions.

(2) Some authors argued that the coronal heating in some stars
including the Sun is probably related to dissipation of currents,
and very strong X-ray emissions arise from variation of magnetic
fields (Galsgaard \& Parnell 2004; Peter et al. 2004). Analogously,
if the corona exists above the disc in MC-II model, we expect that
the corona above $r_{out}$ might be heated by the induced current
due to the screw instability of the non-axisymmetric magnetic field.
Thus the ratio of the disc bolometric flux to the power-law flux is
greater for the inner hotspot than that for the outer hotspot, and
the feature of X-ray radiation given by R02 can be interpreted based
on MC-II model with corona, i.e., there exists a systematic shift
away from disc dominated flux with the increasing power-law flux as
the HFQPO pairs shift from the higher frequency to the lower
frequency based on the observations of XTE J1550-564 and GRO
J1655-40.

(3) Since the two hotspots are produced by two different mechanisms,
we can explain the observations that HFQPOs do not always appear in
simultaneous pairs, because the condition for the magnetic torque
and that for the screw instability would not be satisfied at the
same time. Since the inner and outer hotspots are related to two
different mechanisms based on the non-axisymmetric magnetic fields
in the plunging region and the BH horizon, either the fluctuations
of the magnetic fields at different position or the criterion of the
screw instability could affect the HFQPO pairs. Unfortunately, we
cannot discuss this issue in a quantitative way due to the unknown
origin of the magnetic field near the BH.

(4) According to equation (21) the HFQPO frequency is inversely
proportional to the BH mass, which providers a strong limit to the
fits. We can neither fit the HFQPO with 92 Hz in XTE J1550-564 nor
the 3:2 HFQPO pair (67, 41Hz) in GRS 1915+105 based on equation (21)
for the measured BH masses. Furthermore, the 3:2 HFQPO pair (168,
113Hz) of GRS 1915+105 are not fitted, because a systematic shift
away from disc dominated flux with the increasing power-law flux as
the HFQPO pair shift from the higher frequency to the lower
frequency has not been observed in this source.

(5) As argued above, the higher and lower HFQPO frequencies are
related to the inner and outer hotspots rotating with the angular
velocity located at $r_{in}$ and $r_{out}$, respectively. According
to equation (21) the HFQPO frequencies are inversely proportional to
the BH mass, and the spin is related to the HFQPO frequencies via
$r_{in}$ and $r_{out}$, where the HFQPO frequencies are calculated
by the angular velocity. Since the evolution time scale of the BH
mass and spin are much longer than the observation time scale, the
two quantities are considered fixed in calculations.

(6) As pointed out in R02, the 3:2 HFQPO pairs could have a small,
but significant, change in frequency. For example, the HFQPO
frequencies observed in XTE J1550-564 deviate from 274 Hz and 184 Hz
during different outbursts. These small deviations in HFQPO
frequencies can be calculated by resolving \textbf{FIE} with the
adjustable parameters $n$ and $\delta$ as shown in Table 2.


(7) As argued above, both the non-axisymmetric magnetic field
configuration and the corona above the disc surface are required for
interpreting the 3:2 HFQPO pairs observed in the SPL state of the
BHXBs. Thus we can interpret the observation that sometimes the 3:2
HFQPO pairs cannot be observed in the SPL state, if the magnetic
field configuration is axisymmetric. Based on the same reason we
predict that the 3:2 HFQPO pairs could be observed with the very
steep emissivity index, if the magnetic field configuration is
non-axisymmetric. However, the HFQPO pairs cannot be produced with
the emissivity index for the axisymmetric magnetic field
configuration.

(8) According to $B_D \propto r^{-n}$ the ratio of the magnetic
field at $r_{out}$ to that at $r_{in}$ can be written as $R_B \equiv
(B_D)_{out}/(B_D)_{in}=(r_{in}/r_{out})^n$. Based on \textbf{FIE} we
have the curves of $R_B$ for XTE J1650-564 vary with $a_*$ as shown
in Figure 5, from which a strongly constraint to the BH spin can be
found. The BH spin should be very high, or at least greater than
some intermediate value to avoid the magnetic field at $r_{out}$ too
low to produce an outer hotspot.


(9) There is less certain evidence that the same 3:2 ratio occurs
for QPOs observed in low-mass active galactic nuclei, Sgr A$^*$,
(T\"{o}r\"{o}k 2005a, b; Ashenbach 2004a, b) and in a few nearby
Seyferts (Lachowicz et al. 2006). MC-II model can be used to fit
QPOs pairs observed in the super-massive BH systems, and we obtain
the curves of the emissivity index $\beta$, the power-law index $n$
and the parameter $\delta$ varying with the BH spin of for Sgr A$^*$
as shown in Figure 6.


We can find from Figure 6 that the parameter n varies from 4.272 to
7.108 with $a_*$ decreasing from 0.998 to 0.853 for
$m_{BH}=4.4\times 10^6$, while it varies from 5.040 to 10.239 with
$a_*$ decreasing from 0.998 to 0.555 for $m_{BH}=2.6 \times 10^6$.

(10) The parameter $n$ is used to indicate the variation of the
large-scale magnetic field with the disc radius, which is a key
parameter for fitting the 3:2 frequency ratio. Based on \textbf{FIE}
we have the variation of the 3:2 frequency ratio with the parameter
n for the given values of the BH spin as shown in Figure 7.


\textbf{Inspecting Figure 7, we find that the frequency ratio
decreases monotonically with the increasing parameter $n$. The
greater $n$ corresponds to the less ratio for the given spin, and
the greater spin corresponds to both the less value and less
variation of $n$ for the given ratio. For XTE J1550-564, e.g., the
ratio increases about 5.7\% as the value of $n$ decreases about 10\%
for  $a_*=0.5$, while the same variation occurs as $n$ only
decreases about 1.9\% for $a_*=0.998$ . This result implies that the
observed 3:2 frequency ratio is reached for a very narrow range of
$n$ with the greater BH spin.}

(\textbf{11}) Finally, we discuss the values of the parameters
$\alpha_H$ and $\alpha_m$, which are used to indicate the disc
thickness at $r_{in}$ and the strength of the accretion rate,
respectively. The emissivity index of XTE J1550-564 varies with the
BH spin for different values of $\alpha_H$ and $\alpha_m$ are shown
in Figure \textbf{8}.


Inspecting Figure \textbf{8a}, we find that the values of the
emissivity index are insensitive to the values of $\alpha_H$ and
$\alpha_m$, provided that they are limited to $\alpha _H \le 0.1$
and $\alpha_m \le 1.0$. These value ranges are reasonable for thin
accretion disc in MC-II model. In addition, based on the features of
MC-II model, the 3:2 HFQPO pairs are also insensitive to the
parameters with the above limited values.\\\\

\noindent {\bf Acknowledgements:}\quad This work is supported by the
National Natural Science Foundation of China under grant 10873005
and National Basic Research Program of China under grant
2009CB824800. We are grateful to the anonymous referee for his (her)
helpful comments on the manuscript.


\newpage

\begin{figure}
\vspace{0.5cm}
\begin{center}
\includegraphics[width=12.0cm]{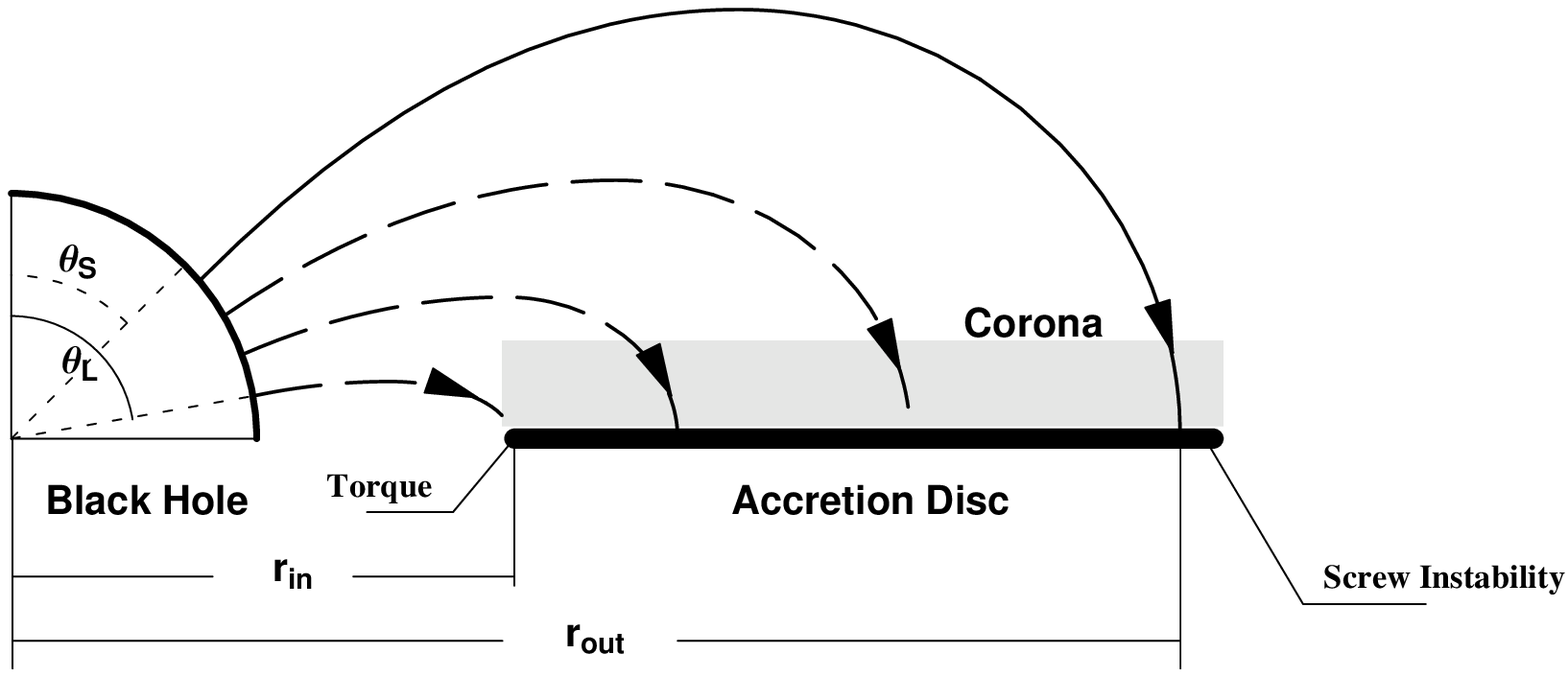}
\caption{The Poloidal magnetic field configuration of MC-II model.}
\label{fig2}
\end{center}
\end{figure}

\begin{figure*}
\vspace{0.5cm}
\begin{center}
{\hfill\hfill
\includegraphics[width=7.5cm]{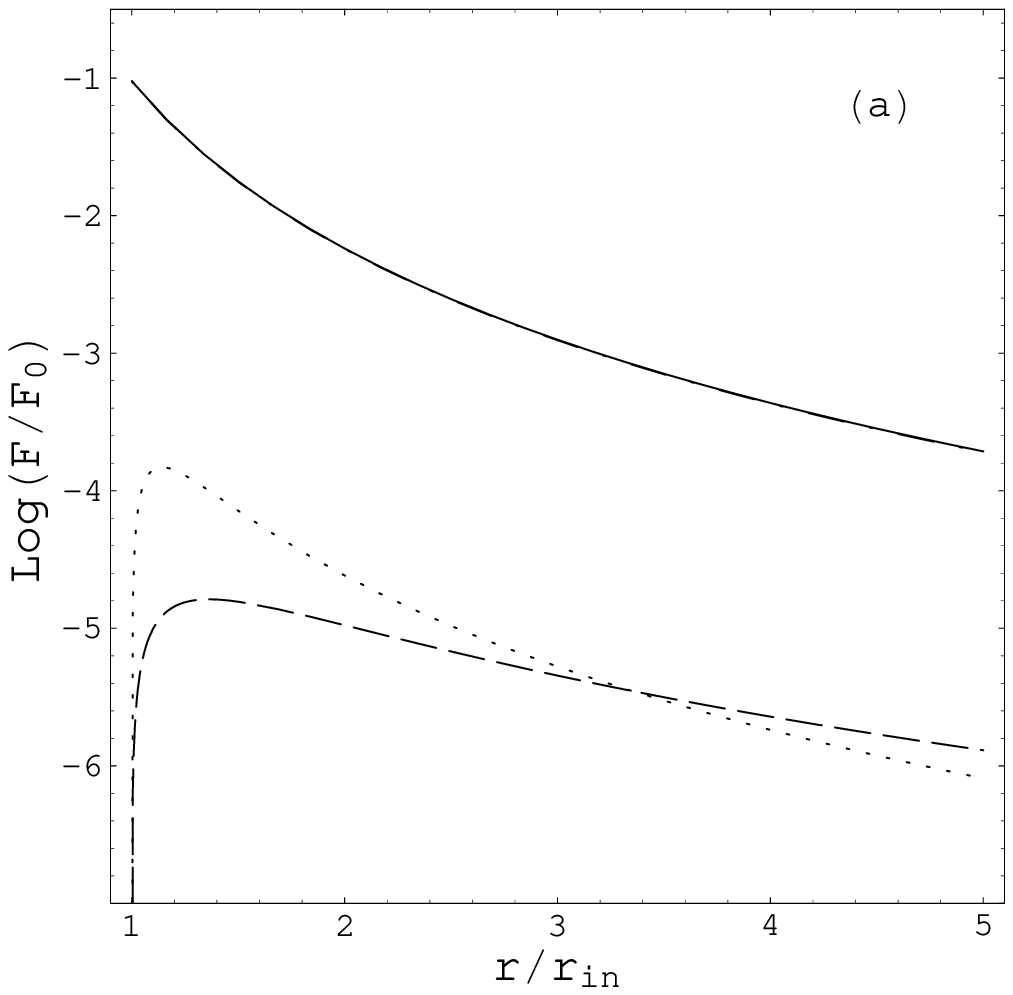}\hfill
\includegraphics[width=7.5cm]{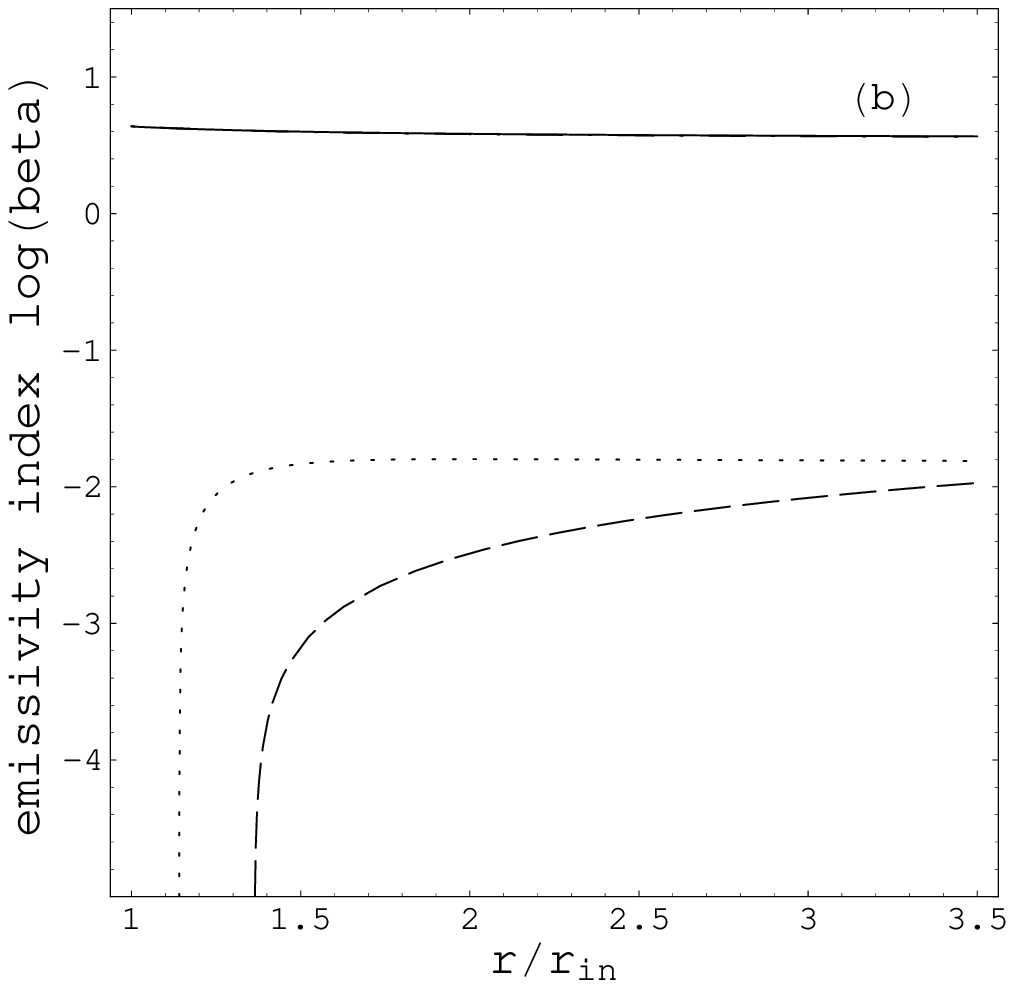}
\hfill\hfill}

\caption{(a) Curves of $\log \left( F_{gin}/F_0 \right)$ , $\log
\left(F_{DA} /F_0  \right)$ and $\log \left( F_{MC} / F_0 \right)$
 versus $r / r_{in}$ in solid, dashed and
dotted lines, respectively. (b) Curves of $log \beta _{gin} $, $log
\beta _{DA} $ and $log \beta _{MC} $ versus $r /r_{in} $ in solid,
dashed and dotted lines, respectively.  These curves are plotted for
XTE J1550-564 with $m_{BH} = 10.8$,  $a_*=0.7$ , $n=9.99$  and
$\delta=3.29 \times 10^{-4}$ . } \label{fig3}
\end{center}
\end{figure*}

\begin{figure*}
\vspace{0.5cm}
\begin{center}

{\hfill\hfill
\includegraphics[width=5.4cm]{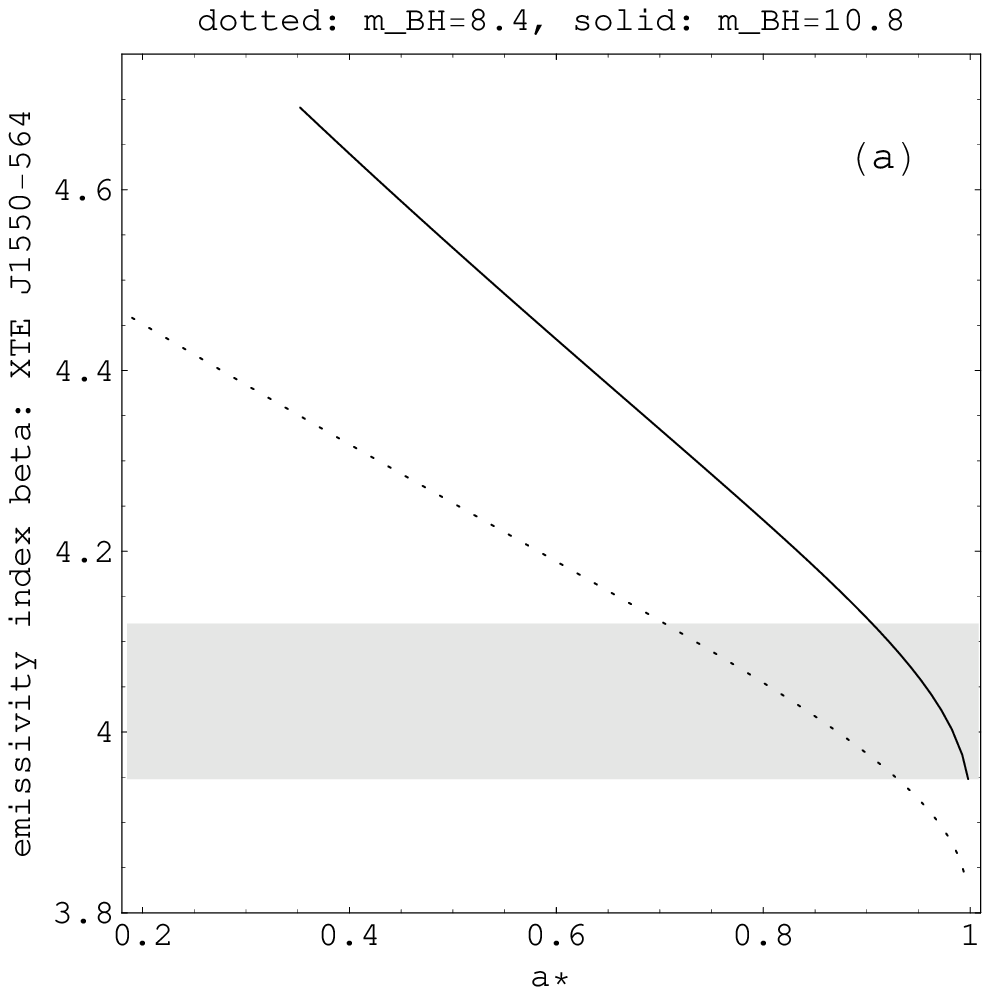}\hfill
\includegraphics[width=5.4cm]{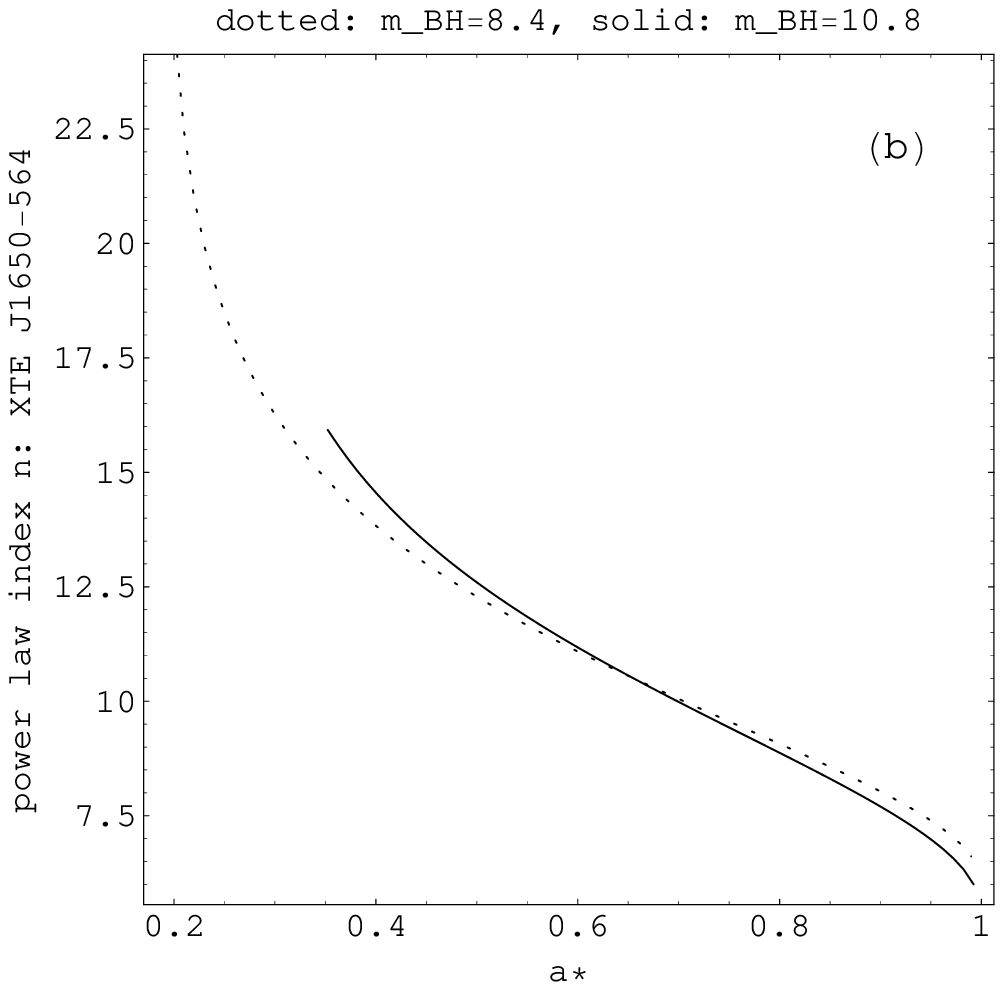}\hfill
\includegraphics[width=5.4cm]{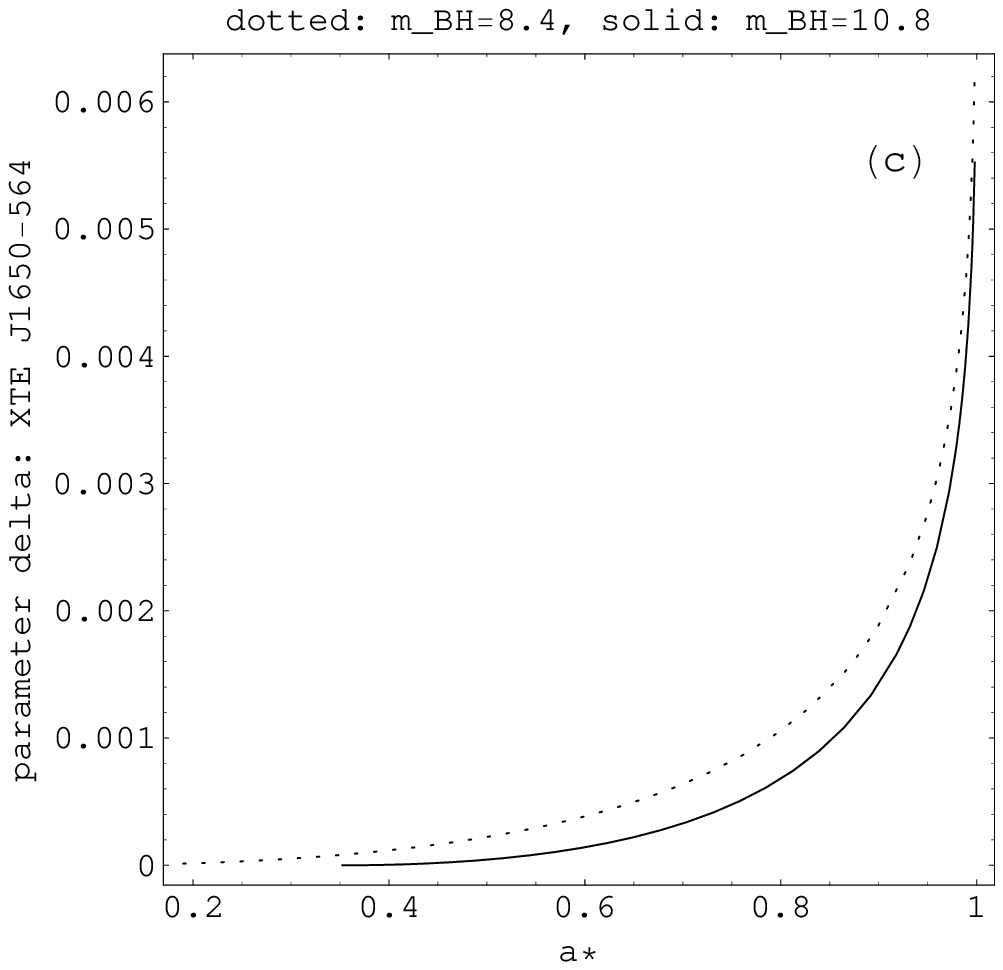}
\hfill\hfill}

\caption{The curves of the emissivity index $\beta$, the power-law
index $n$ and the parameter $\delta$ versus the black hole spin
$a_*$ for XTE J1550-564 are shown in Figures 3a, 3b and 3c,
respectively. The solid and dotted lines correspond to the upper and
lower limits to the black hole mass, respectively. The shaded region
in Fig. 3a indicates $3.95 < \beta <4.12$.}

\label{fig4}
\end{center}
\end{figure*}

\begin{figure*}
\vspace{0.5cm}
\begin{center}

{\hfill\hfill
\includegraphics[width=5.4cm]{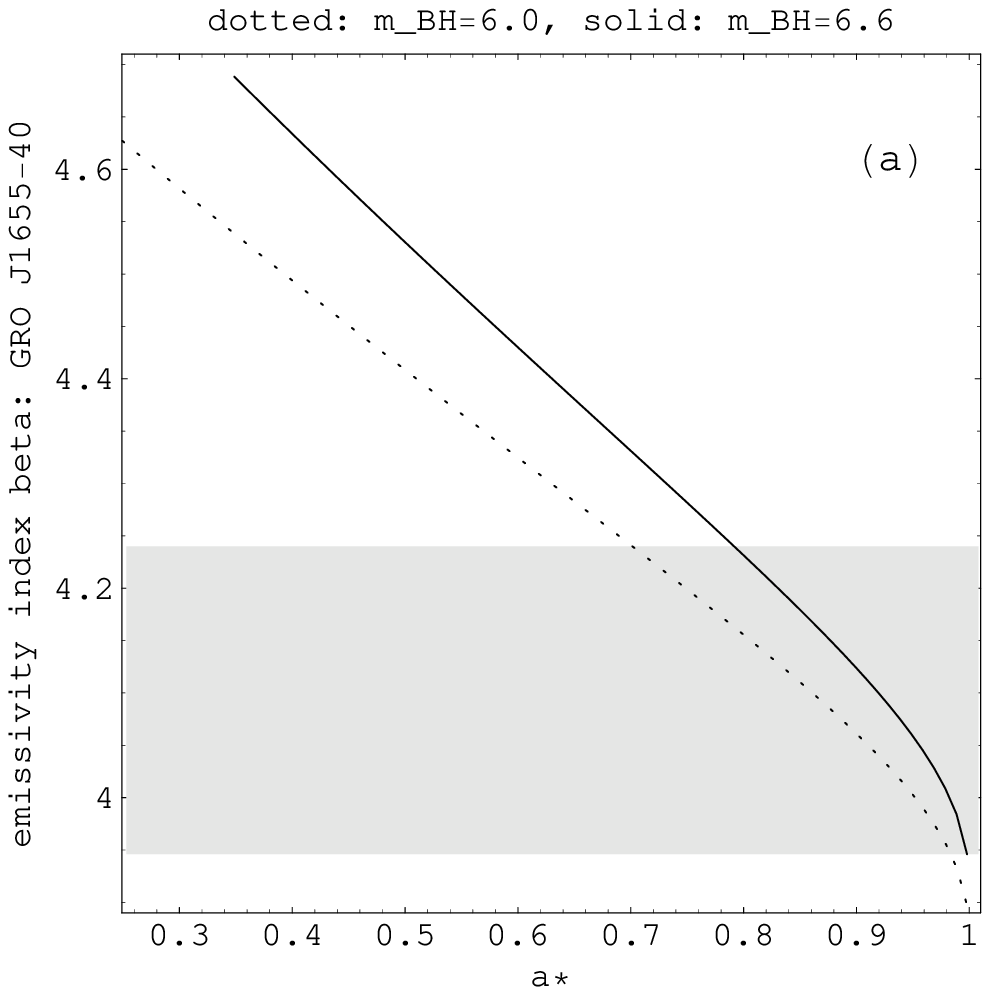}\hfill
\includegraphics[width=5.4cm]{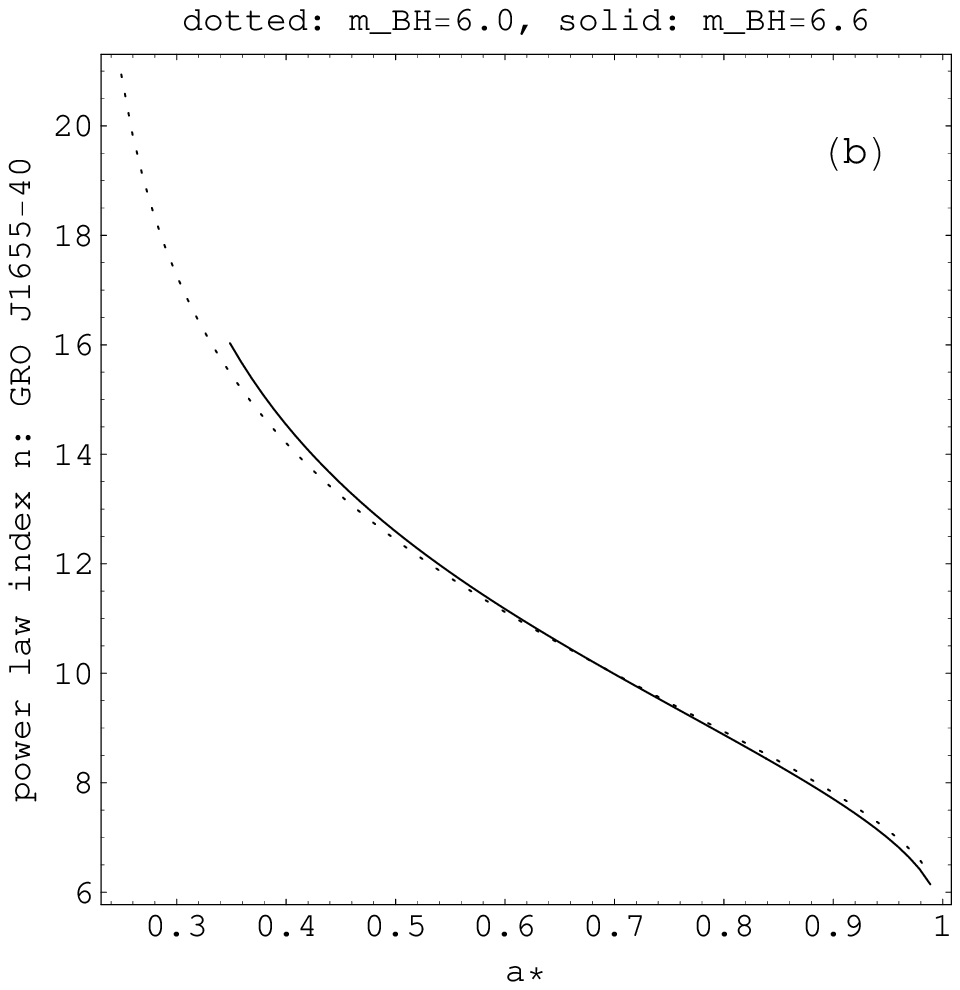}\hfill
\includegraphics[width=5.4cm]{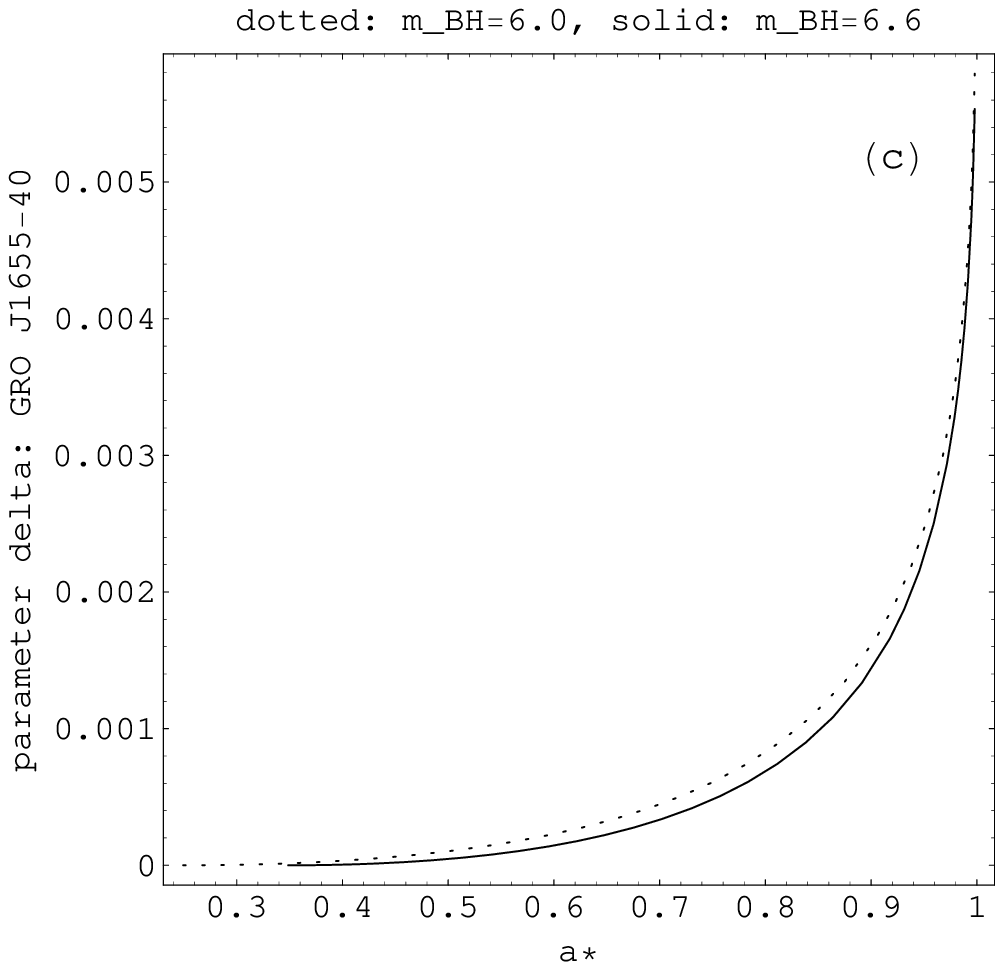}
\hfill\hfill}

\caption{ The corresponding curves for GRO J1655-40 with the same
caption as given in Fig.3, except that the shaded region indicates
 $3.95 < \beta <4.24$.}

\label{fig4}
\end{center}
\end{figure*}

\begin{table}
\label{tab1}
\caption{Calculation results for the 3:2 HFQPO pairs and the
emissivity index based on MC-II model.}
\begin{center}
\begin{tabular}{ccccccccccc}

\hline \hline&&&&&&&\\
&&{Input Quantities}&&&&&&&{Output Quantities}\\
\cline{2-5} \cline{8-11} {Source}&&&&&&&\\
&{$\nu_{higher}$}&{$\nu_{lower}$}&{$a_*$}&{$m_{BH}$}&&
&&{$n$}&{$\delta$}&{$\beta $}\\
\hline &&&&&&&\\
&      &      &0.5   &8.4   &&&&12.29      &$2.22\times 10^{-4}  $     &4.25\\
&      &      &      &10.8  &&&&12.60      &$4.42\times 10^{-5}  $    &4.54\\
\cline{4-11} &&&&&&&\\
XTE J1550-564
&276Hz &184Hz &0.7   &8.4   &&&&10.05      &$6.37\times 10^{-4}  $     &4.12\\
&      &      &      &10.8  &&&&9.99       &$3.29\times 10^{-4}  $     &4.33\\
\cline{4-11} &&&&&&&\\
&      &      &\ \ \ \ 0.998\ \ \ \
                     &8.4   &&&&6.23       &$6.22\times 10^{-3} $     &3.83\\
&      &      &      &10.8  &&&&5.65       &$5.52\times 10^{-3} $     &3.95\\
\hline &&&&&&&\\

&      &      &0.5   &6.0  &&&&12.44      &$1.03\times 10^{-4}  $     &4.41\\
&      &      &      &6.6  &&&&12.59      &$4.60\times 10^{-5}  $     &4.53\\
\cline{4-11} &&&&&&&\\
GRO J1655-40
&450Hz &300Hz &0.7   &6.0  &&&&9.99       &$4.45\times 10^{-4}  $     &4.24\\
&      &      &      &6.6  &&&&9.99       &$3.33\times 10^{-4}  $     &4.33\\
\cline{4-11} &&&&&&&\\
&      &      &0.998 &6.0  &&&&5.87       &$5.79\times 10^{-3}  $     &3.90\\
&      &      &      &6.6  &&&&5.66       &$5.53\times 10^{-3}  $     &3.95\\
\hline
\end{tabular}
\end{center}
Notes: {The input quantities of XTE J1550-564 and GRO J1655-40 are
taken from RM06. The parameter $\delta$  is defined in equation
(\ref{eq20}) as the fraction of the power of the magnetic torque to
keep the inner edge at $r_{in}$.}\\

\end{table}

\begin{table}
\label{tab2}
\caption{Fits of the small changes in HFQPO frequency of XTE
J1650-564 with $m_{BH}=10.8$.}
\begin{center}
\begin{tabular}{ccccccccccc}
\hline \hline\\
&{Source}&&&{1998-99A}&{1998-99B}&{2000 Apr 30-May 9}\\
\hline\\
&{$\nu_{higher}$(Hz)}&&&$281.7\pm 1.5$ &277.7 &$269.4\pm2.7$\\
\\
&{$\nu_{lower}$(Hz)}&&&187.8          &$185.1\pm3.5$ &178.6\\
\\
& \textbf{Ratio} &&&$1.5^{+0.01}_{-0.01}$ &$1.5^{-0.03}_{+0.03}$
&$1.5^{+0.02}_{-0.01}$ \\
\hline &\\
&&$\delta$ &&$3.46^{-0.23}_{+0.24} \times 10^{-5}$ &$4.12 \times
10^{-5}$  &$5.68^{-0.53}_{+0.56}\times
10^{-5}$\\\\
&0.5&$n$ &&$12.64^{-0.14}_{+0.14}$  &$12.60^{+0.57}_{-0.52}$
&$12.40^{-0.25}_{+0.27}$\\\\
&&$\beta$ &&$4.565^{+0.008}_{-0.008}$  &$4.544$    &$4.502^{+0.014}_{-0.013}$\\
\cline{2-7} &\\
 &&$\delta$ &&$3.06^{-0.06}_{+0.06}\times 10^{-4}$ &$3.22\times 10^{-4}$   &$3.57^{-0.12}_{+0.11}\times
 10^{-4}$\\\\
$a_{*}$&0.7&$n$ &&$9.99^{-0.10}_{+0.10}$   &$9.98^{+0.39}_{-0.35}$
&$9.88^{-0.18}_{+0.19}$\\\\
&&$\beta$ &&$4.356^{+0.006}_{-0.005}$  &$4.341$    &$4.310^{+0.010}_{-0.009}$\\
\cline{2-7} &\\
&&$\delta$&&$5.47^{-0.01}_{+0.01}\times 10^{-3}$     &$5.51\times
10^{-3}$   &$5.59^{-0.04}_{+0.03}\times
10^{-3}$\\\\
&0.998&$n$      &&$5.61^{-0.02}_{+0.03}$  & $5.64^{+0.05}_{-0.04}$
&$5.69^{-0.04}_{+0.05}$\\\\
&&$\beta$ &&$3.959^{+0.003}_{-0.003}$  &$3.951$    &$3.935^{+0.005}_{-0.005}$\\
\hline

\end{tabular}
\end{center}
Notes: {(1) The super- and sub-scripts in $\delta$, $n$ and $\beta$
correspond to up and down deviations of HFQPO frequencies,
respectively. (2) The value of $\delta$ only depends on $a_{*}$  and
$\nu_{higher}$ , being independent of the deviation of $\nu_{lower}$
in the case of 1998-99B.}\\

\end{table}

\begin{figure*}
\vspace{0.5cm}
\begin{center}
{\hfill\hfill
\includegraphics[width=7.5cm]{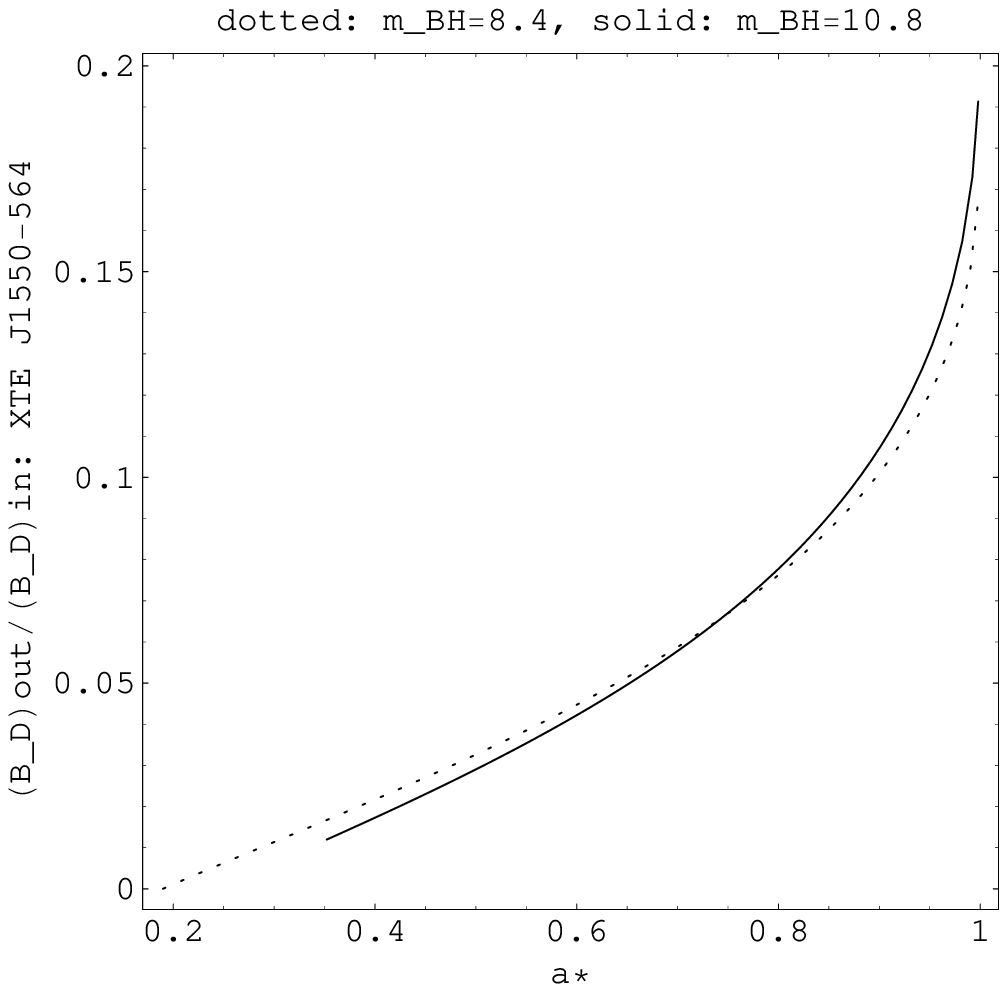}\hfill
\includegraphics[width=7.5cm]{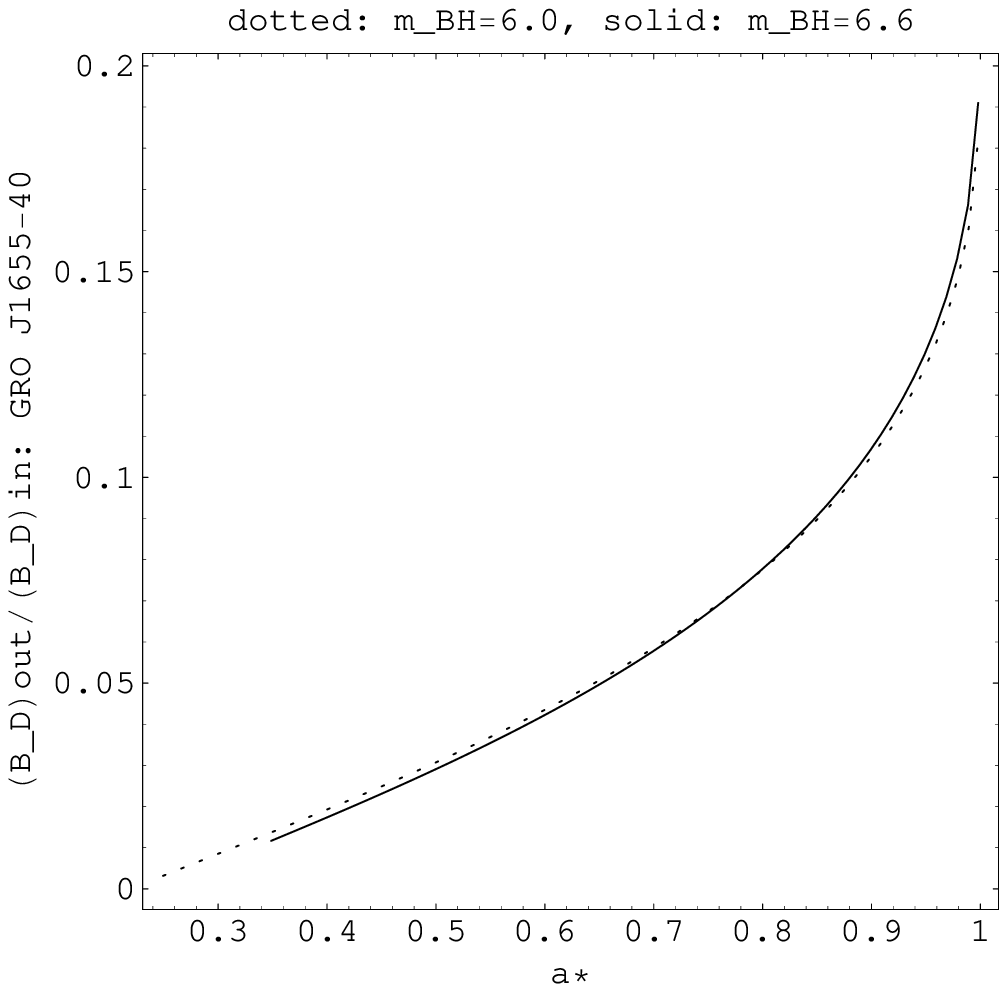}
\hfill\hfill}

\caption{The curves of the ratio  $R_B$ versus the black hole spin
$a_*$  for (a) XTE J1650-564 and for (b) GRO J1655-40. The solid and
dotted lines correspond to the higher and lower limits to the black
hole mass, respectively.}

\label{fig5}
\end{center}
\end{figure*}

\begin{figure*}
\vspace{0.5cm}
\begin{center}

{\hfill\hfill
\includegraphics[width=5.4cm]{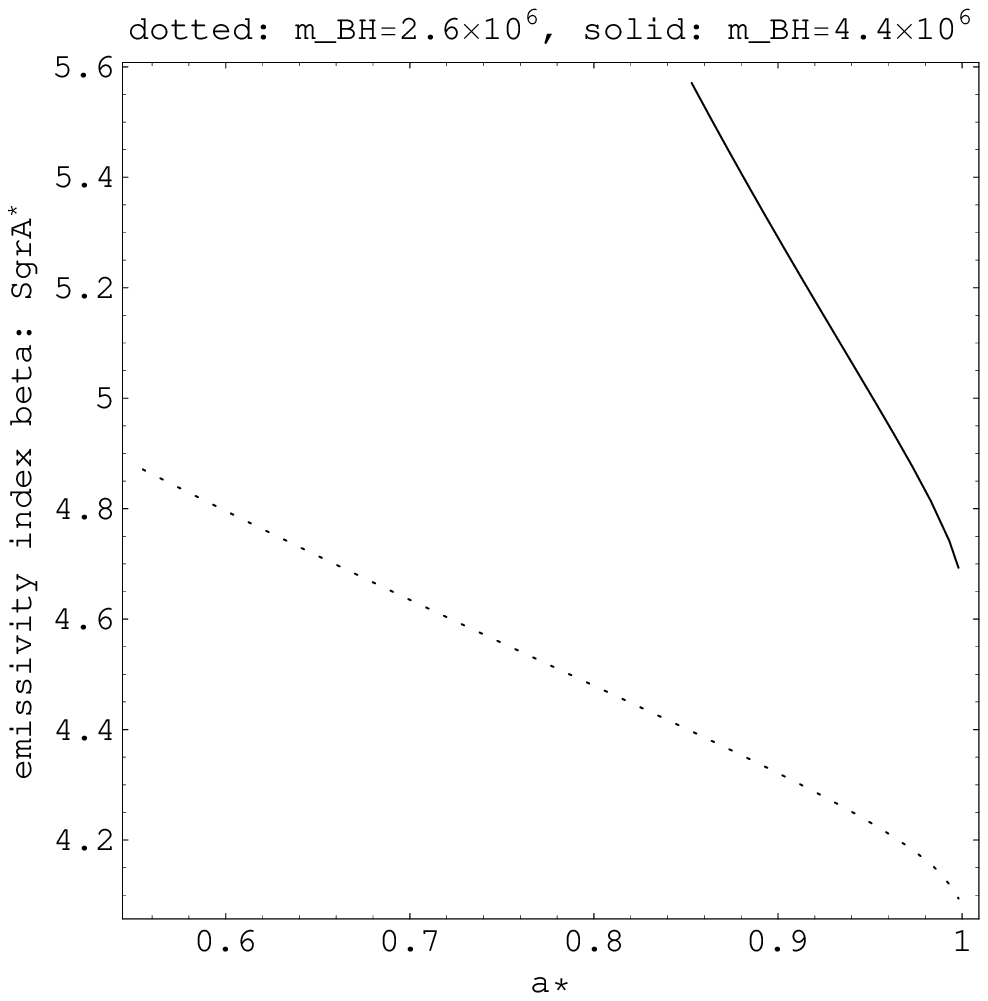}\hfill
\includegraphics[width=5.4cm]{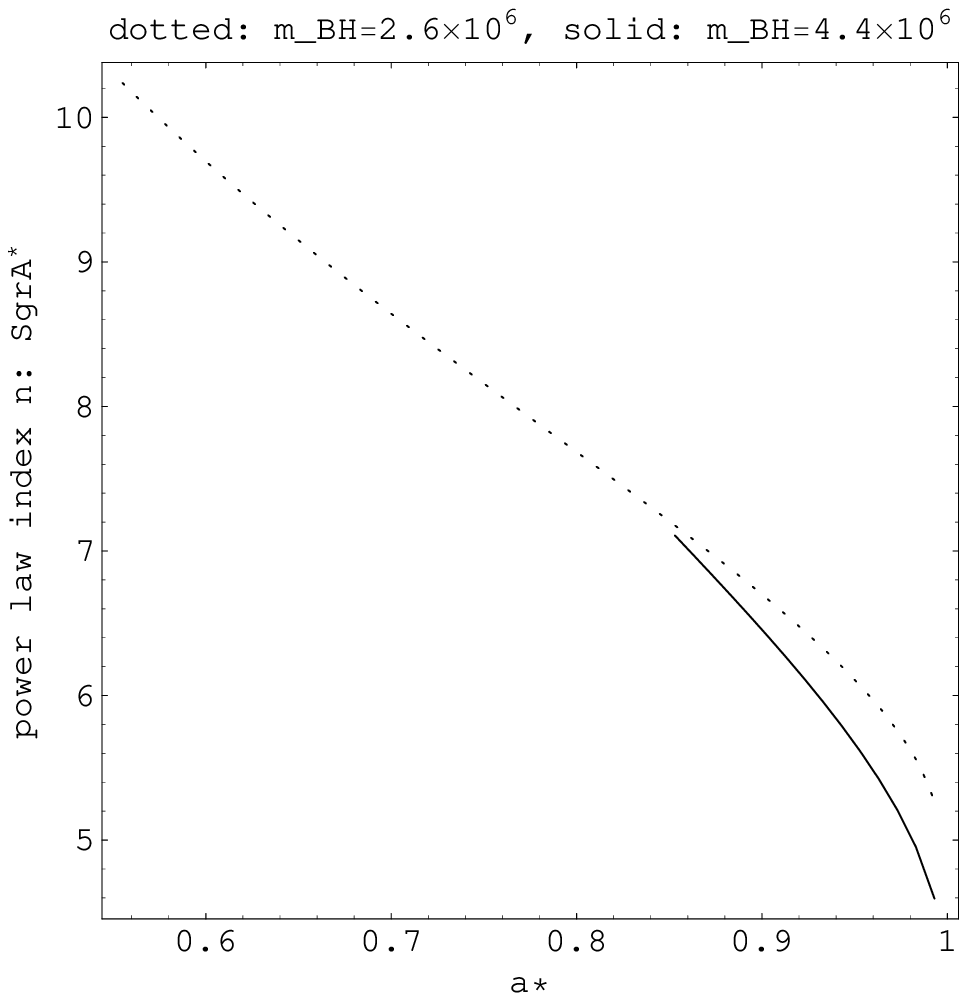}\hfill
\includegraphics[width=5.4cm]{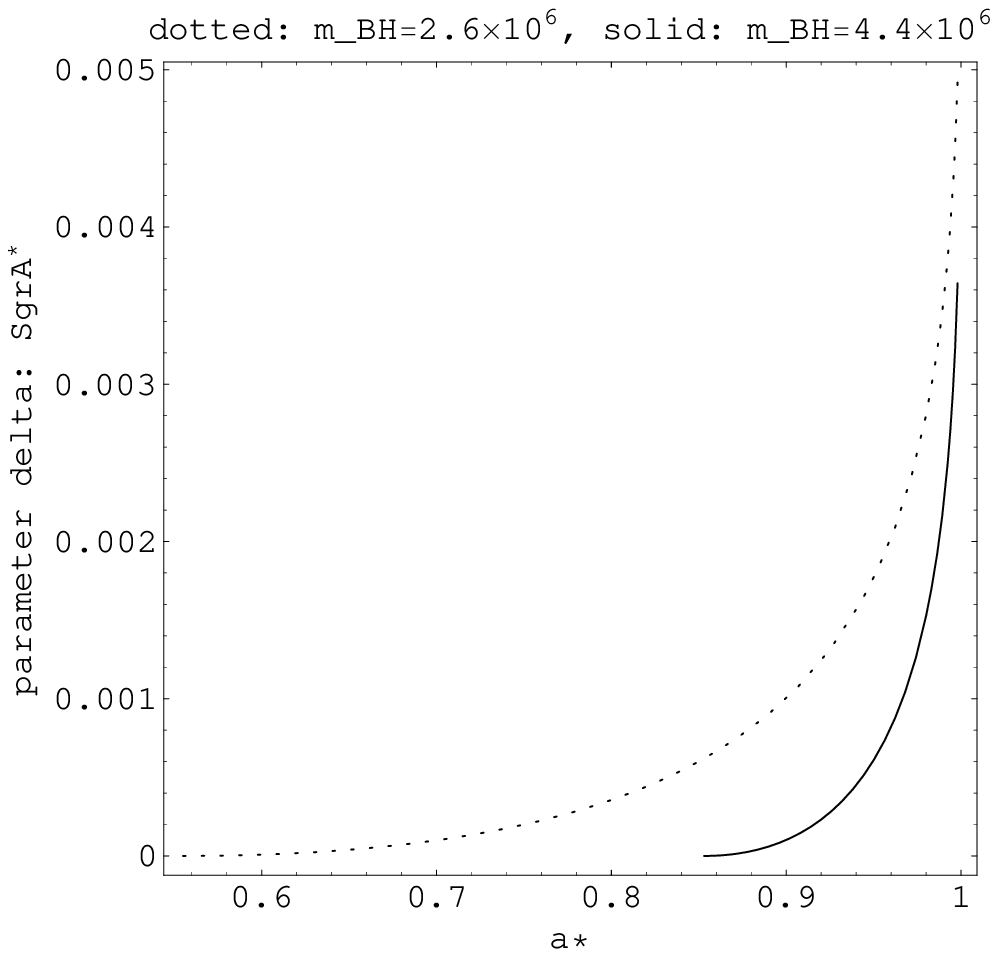}
\hfill\hfill}

\caption{ The curves of the emissivity index $\beta$, the power-law
index $n$ and the parameter $\delta$  versus the black hole spin
$a_*$  for Sgr A* are shown in Figures 6a, 6b and 6c, respectively.
The solid and dotted lines correspond to the higher and lower limits
to the black hole mass ($m_{BH}=4.4 \times 10^6$  and $m_{BH}=2.6
\times 10^6$ ).}

\label{fig6}
\end{center}
\end{figure*}

\begin{figure*}
\vspace{0.5cm}
\begin{center}
{\hfill\hfill
\includegraphics[width=7.5cm]{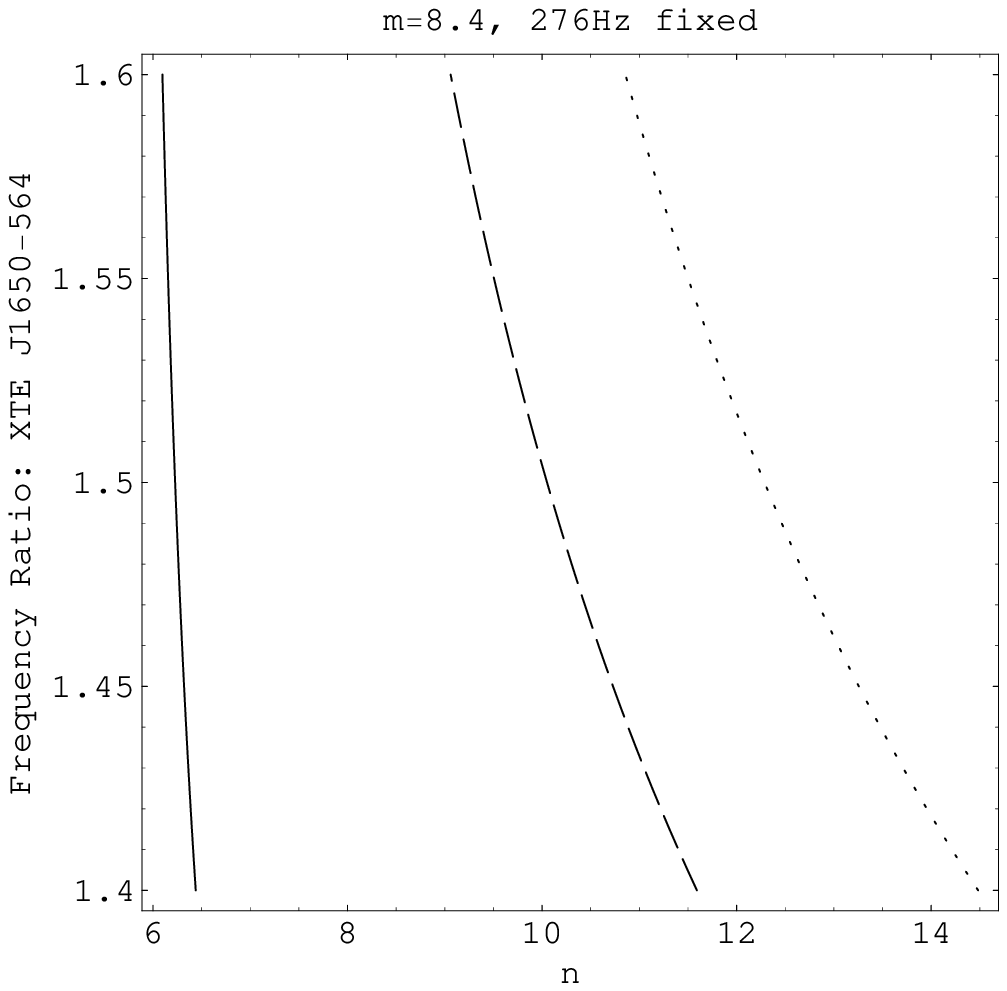}\hfill
\includegraphics[width=7.5cm]{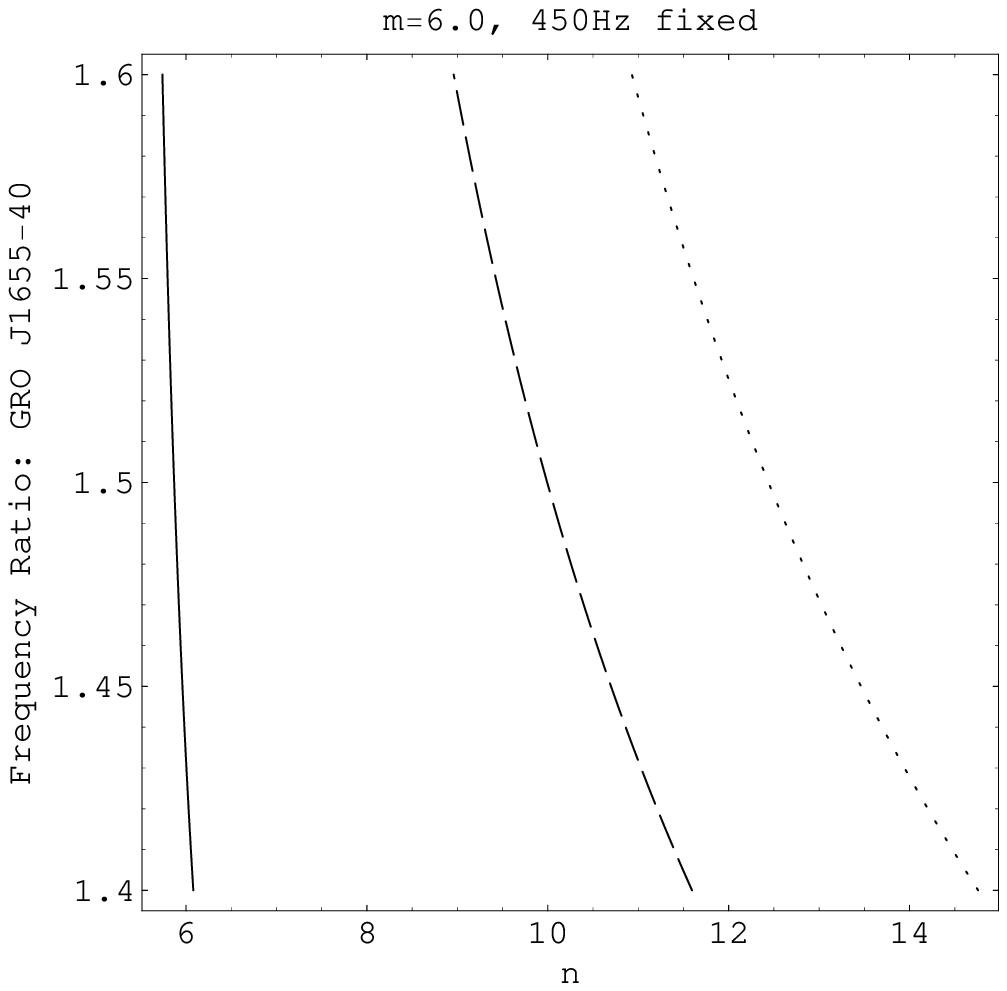}
\hfill\hfill}

\caption{The variation of the frequency ratio with the parameter n
for (a) XTE J1550-564 and for (b) GRO J1655-40. The solid, dashed
and dotted lines correspond to $a_*=0.998$, $0.7$ and $0.5$,
respectively.}

\label{fig7}
\end{center}
\end{figure*}

\begin{figure*}
\vspace{0.5cm}
\begin{center}
{\hfill\hfill
\includegraphics[width=7.5cm]{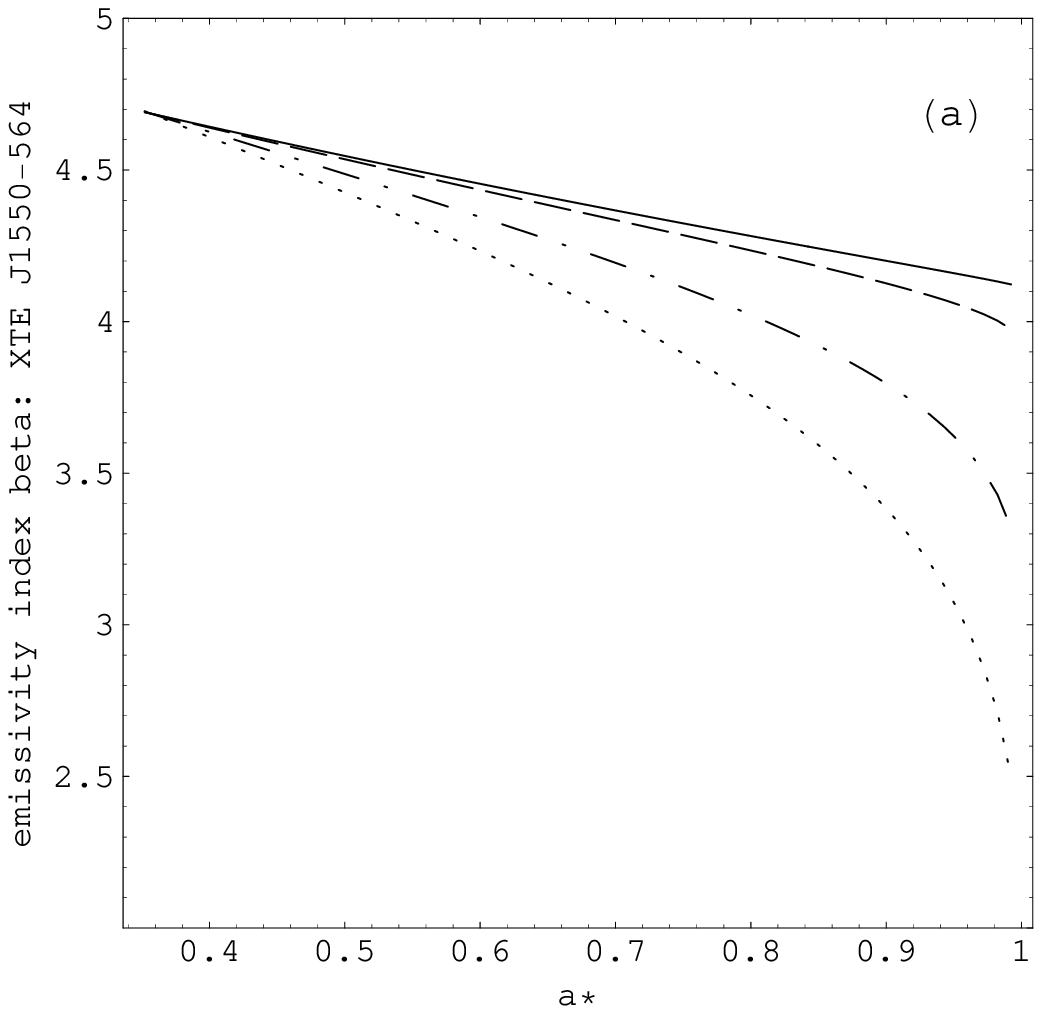}\hfill
\includegraphics[width=7.5cm]{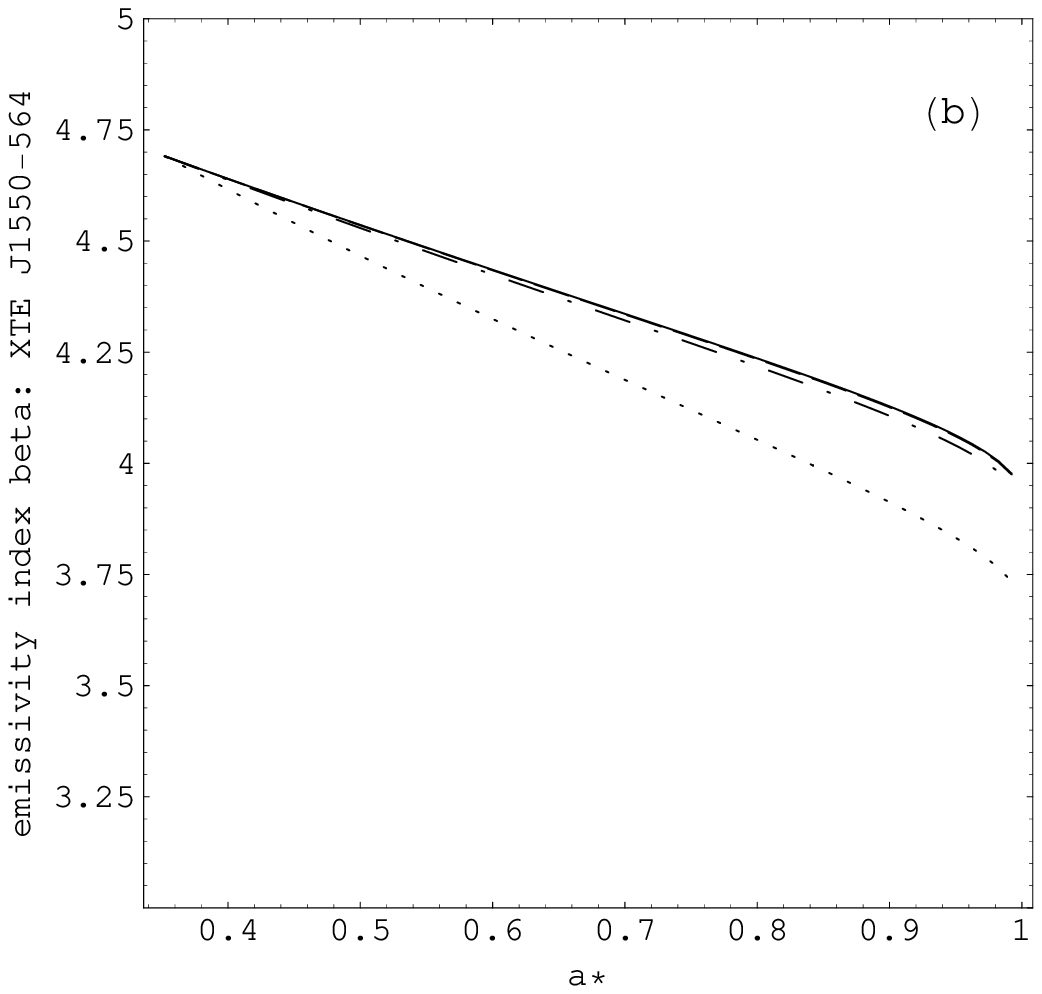}
\hfill\hfill}

\caption{The curves of emissivity index of XTE J1550-564 versus the
black hole spin for (a) $\alpha_m=0.1$, and $\alpha_H=0.01$, 0.1,
0.5 and 1.0 in solid, dashed, dot-dashed and dotted lines,
respectively, and for (b) $\alpha_H=0.1$, and $\alpha_m=0.01$, 0.1,
1.0 and 10.0 in solid, dashed, dot-dashed and dotted lines,
respectively. The black hole mass is taken as $m_{BH}=10.8$. }

\label{fig8}
\end{center}
\end{figure*}

\end{document}